\PassOptionsToPackage{table,xcdraw}{xcolor}
\documentclass[screen, acmsmall]{acmart}

\usepackage[table,xcdraw]{xcolor}
\usepackage{algorithmic}
\usepackage{textcomp}
\usepackage{multirow}
\usepackage{adjustbox}
\usepackage{pifont}
\usepackage{amsthm,amsmath,amsfonts} 
\usepackage{url}
\usepackage{threeparttable}
\usepackage{booktabs}
\usepackage{bbding}
\usepackage{tablefootnote}
\usepackage{hyperref}
\usepackage{tcolorbox}
\usepackage{enumitem}
\usepackage{stfloats}
\usepackage{graphicx}
\usepackage{subcaption} 

\AtBeginDocument{%
  }


\setcopyright{rightsretained}
\acmDOI{10.1145/3689794}
\acmYear{2024}
\acmJournal{PACMPL}
\acmVolume{8}
\acmNumber{OOPSLA2}
\acmArticle{354}
\acmMonth{10}
\acmSubmissionID{oopslab24main-p812-p}
\received{2024-04-06}
\received[accepted]{2024-08-18}

\begin{document}

\makeatletter
\makeatother

\newcommand{\tool}{tool}
\newcommand{\ngram}{Ngram-NT}
\newcommand{\tech}{DAN}
\newcommand{\old}{Ngram-NT}
\newcommand{\oldvariant}{CSN}
\newcommand{\codebert}{CodeBERT-NT}

\newcommand\note[1]{\textit{\textcolor{blue!80}{[yc: #1]}}}
\newcommand\jj[1]{\textit{\textcolor{orange}{[Junjie: #1]}}}
\newcommand\jun[1]{\textit{\textcolor{magenta}{[Jiang: #1]}}}
\newcommand\todo[1]{\textit{\textcolor{red}{[#1]}}}

\newcommand{\revise}[1]{#1} 

\newcommand{\del}[1]{} 

\definecolor{top3}{gray}{0.75}

\title{Dependency-Aware Code Naturalness}

\author{Chen Yang}
\orcid{0000-0003-0759-940X}
\affiliation{%
  \institution{College of Intelligence and Computing, Tianjin University}
  \city{Tianjin}
  \country{China}
}
\email{yangchenyc@tju.edu.cn}

\author{Junjie Chen}
\authornote{Corresponding author}
\orcid{0000-0003-3056-9962}
\affiliation{%
  \institution{College of Intelligence and Computing, Tianjin University}
  \city{Tianjin}
  \country{China}
}
\email{junjiechen@tju.edu.cn}

\author{Jiajun Jiang}
\orcid{0000-0003-1983-6572}
\affiliation{%
  \institution{College of Intelligence and Computing, Tianjin University}
  \city{Tianjin}
  \country{China}
}
\email{jiangjiajun@tju.edu.cn}

\author{Yuliang Huang}
\orcid{0009-0005-3840-4988}
\affiliation{%
  \institution{College of Intelligence and Computing, Tianjin University}
  \city{Tianjin}
  \country{China}
}
\email{huangyuliang@tju.edu.cn}

\begin{abstract}
Code naturalness, which captures repetitiveness and predictability in programming languages, has proven valuable for various code-related tasks in software engineering. 
However, precisely measuring code naturalness remains a fundamental challenge. 
Existing methods measure code naturalness over individual lines of code while ignoring the deep semantic relations among different lines, e.g., program dependency, which may negatively affect the precision of the measure.
Despite the intuitive appeal of extending the code naturalness measure to the code dependency domain (as there are some work that have initiated the utilization of code dependency for diverse code-related tasks), this assumption remains unexplored and warrants direct investigation.
In this study, we aim to perform the first empirical study to investigate whether incorporating code dependency, instead of analyzing individual lines, can enhance the precision of measuring code naturalness.

To achieve that, we first propose a new method named \tech{} for measuring code naturalness by incorporating the rich dependency information in the code. Specifically, \tech{} extracts multiple sequences of code lines by traversing the program dependency graph, where different code lines are connected by dependencies in each sequence, and then the code naturalness will be measured by taking each sequence as a whole. In this way, the dependency information can be well captured.
Finally, we have conducted an extensive study to evaluate the influence of code dependency for measuring code naturalness with \tech{}, and compared it with the state-of-the-art methods under three emerging application scenarios of code naturalness. The results demonstrate that \tech{} can not only better distinguish natural and unnatural code, but also substantially boost two important downstream applications of code naturalness, i.e., distinguishing buggy and non-buggy code lines and data cleansing for training better code models, reflecting the significance of code dependency in measuring code naturalness.

\end{abstract}

\maketitle

\section{Introduction}
\label{sec:introduction}

Natural languages (NL) exhibit notable repetitiveness and predictability, which facilitates reliable and efficient communication among humans ~\cite{pilkington1996language, buggy_code_naturalness}. 
This property, commonly referred to as \textit{naturalness}, has been effectively harnessed in the area of Natural Language Processing (NLP)~\cite{kowsari2019text, cavnar1994n}. 
Inspired by it, programming languages, as a special form of human-oriented languages, have been also demonstrated repetitive and predictable~\cite{buggy_revisited, software_naturalness}.
This property of code has been successfully exploited to facilitate various code-related tasks, such as code generation ~\cite{natgen}, code completion~\cite{2014Completion}, and code de-obfuscation~\cite{raychev2015predicting}.
It also constitutes the foundation for the rapid development of statistical language modeling of source code~\cite{zheng2023codegeex, wang2021codet5}.

Due to the importance of this property, how to measure the naturalness of code is one of the most fundamental tasks in this area.
In the literature, some methods of measuring code naturalness have been proposed~\cite{software_naturalness,tu2014localness,buggy_code_naturalness, buggy_revisited}.
\revise{
For example, Hindle et al.~\cite{software_naturalness} applied the n-gram model to measure cross entropy of source code as code naturalness.
As source code tends to be more repetitive within a small module, Tu et al.~\cite{tu2014localness} further incorporated a cache model to exploit localness of source code for measuring code naturalness more precisely.
} 
Then, Ray et al.~\cite{buggy_code_naturalness} 
proposed to normalize code naturalness according to statement types in order to balance the naturalness of different syntax types.
Further, to relieve the limitations of the n-gram model (e.g., trained on small datasets and thus inapplicable to cross-project scenarios), Khanfir et al.~\cite{khanfir2022codebert-nt} proposed \codebert{} to use the pre-trained language model, i.e., CodeBERT~\cite{feng2020codebert} instead of n-gram, to measure code naturalness based on the prediction confidence on masked tokens in code.

Although code naturalness can be measured more precisely with more methods being proposed, almost all of them share the same workflow, i.e., measuring naturalness of each line in a complete code snippet (e.g., a method or a file) individually.
However, humans and compilers tend not to understand a single line without considering sufficient contexts, especially the lines of code that are depended on~\cite{rahman2019natural_revisited,busjahn2015eye}.
Hence, such a way isolating each code line could negatively affect the measure of code naturalness.

In this work, we aim to perform the first empirical study to investigate whether incorporating code dependency, instead of isolating each line, can improve the precision of measuring code naturalness.
Intuitively, considering all lines in the complete code snippet together in a sequential manner can be a straightforward method to include all dependency information within the code snippet.
However, such a coarse-grained strategy may incur much noise because sequential lines in code may be independent with each other, which has been confirmed in our study (Section~\ref{sec:evaluation}).
Indeed, humans and compilers do not process code in a completely sequential manner~\cite{busjahn2015eye, rahman2019natural_revisited}.

To investigate the influence of code dependency well, we first propose a more reasonable way of incorporating dependency information among lines in code for measuring code naturalness.
We call it \textit{dependency-aware code naturalness} (abbreviated as \tech{}).
Specifically, \tech{} represents the complete code as a program dependency graph (PDG) by performing control- and data- flow analysis.
Then, it traverses all paths in the graph in a DFS (depth first search) manner, each of which can be treated as a sequence of lines with control or data dependency.
\tech{} measures the naturalness of each sequence with some statistical model, which is helpful to incorporate dependency information without incurring noise, and then obtains the naturalness of the complete code snippet by aggregating the naturalness of all these sequences.
Note that our novelty does not lie in designing new statistical models for measuring code naturalness.
In fact, the idea of \tech{} is general and can be applied independently of the used statistical model.
To demonstrate the generalizability of our \tech{} idea, we create two instantiations of \tech{} by integrating the two state-of-the-art statistical models that have been used to measure code naturalness (i.e., the cached n-gram model and the pre-trained CodeBERT model) for evaluation, respectively.
\tech{} can also integrate more advanced statistical models for better performance (even though they were not investigated to measure code naturalness), which will be discussed in Section~\ref{sec:llms}.
For ease of presentation, we call them \textbf{\tech{}$_{\textit{ngram}}$} and \textbf{\tech{}$_{\textit{codebert}}$}, respectively.

Recent work has initiated the utilization of code dependency for diverse code-related tasks~\cite{suo2024mlir, guo2020graphcodebert, yang2024enhancing}, exemplified by GraphCodeBERT~\cite{guo2020graphcodebert}, which leverages program analysis to augment vectorization of code snippets. 
Despite the intuitive appeal of extending the naturalness measure to the code dependency domain, this assumption remains unexplored and warrants direct investigation.
Therefore, despite conceptually simple, \tech{} is the first attempt to incorporate code dependency in measuring code naturalness.
In this way, code semantic information can be effectively utilized to measure code naturalness to some degree.

With \tech{}, we conducted the extensive study to investigate the influence of code dependency on measuring code naturalness compared to the state-of-the-art n-gram-based method~\cite{buggy_code_naturalness} (called \old{} in our work) and \codebert{}~\cite{khanfir2022codebert-nt}, both of which work on a per-line basis for measuring code naturalness.
Specifically, we performed comparisons (\tech{}$_{\textit{ngram}}$ v.s. \old{} and \tech{}$_{\textit{codebert}}$ v.s. \codebert{}, each pair of methods used the same statistical model for investigating the contribution of code dependency more clearly) in three scenarios by carefully designing corresponding experiments.
First of all, we investigated the ability of \tech{} to distinguish natural and unnatural code, which is the core task of measuring code naturalness, by intentionally transforming each piece of natural code in the widely-used HumanEval-X dataset~\cite{zheng2023codegeex} into unnatural code based on a set of transformation rules.
The results show that \tech{} can distinguish natural code and unnatural code better than \old{}/\codebert{} by achieving 41.82\% and 13.41 times more significant difference in the measured naturalness scores between them on average.

We then evaluated \tech{} in two downstream applications of code naturalness: 
(1) distinguishing buggy and non-buggy code lines following the existing work~\cite{buggy_code_naturalness, buggy_revisited, jit}, and 
(2) cleansing training data for building better code generation models, which can generate more natural code without damaging the functionality correctness of generated code, as indicated by the existing work~\cite{natgen, msra_survey}.
For the former, we used all the three datasets that have been used in this application, i.e., Defects4J~\cite{just2014defects4j}, GrowingBugs~\cite{growing}, SmartSHARK~\cite{Herbold2020smartshark}.
The results show that \tech{} helps
identify 26.67\%, 14.81\%, 28.48\% more buggy lines within the top20\% prioritized lines than \old{} on the three datasets respectively, 
and helps identify 32.00\%, 14.29\%, and 40.24\% more buggy lines than \codebert{} accordingly.
For the latter, we used two widely-studied code generation models (i.e., CodeGen-Multi~\cite{nijkamp2022codegen} and GPT-2~\cite{radford2019language})
as the target models due to the model availability and fine-tuning cost, and used the APPS~\cite{hendrycks2021apps} and HumanEval-X~\cite{zheng2023codegeex} datasets since they have been widely used in the area of code generation.
The results show that the training data selected by \tech{} builds better code generation models than the entire training set and the data selected by \old{} and \codebert{} respectively.
Specifically, the code generation models with \tech{} achieve 18.08\% and 16.25\% higher CodeBLEU scores than those with \old{} and \codebert{} on average.
The results demonstrate that the former can generate code that has high textual similarity with the ground-truth (human-written) code, indicating the high readability of generated code to some degree.
In particular, the former even achieves higher code generation accuracy (measured through test execution on generated code) than the latter two.
Overall, the results demonstrate the contribution of incorporating code dependency to improve the precision of measuring code naturalness as well as boost the two important downstream applications.

To sum up, our work makes the following major contributions:
\begin{itemize}

\item We are the first to empirically investigate the influence of code dependency on measuring code naturalness, compared to the state-of-the-art practice of isolating each line in code.

\item We design dependency-aware code naturalness (i.e., \tech{}), a simple but effective method to measure code naturalness by incorporating code dependency among lines.

\item We conducted the extensive study with \tech{} in three scenarios, including the core task of distinguishing natural and unnatural code and two downstream applications of code naturalness, demonstrating its significant contribution in all the three scenarios.

\end{itemize}

\section{Background}
\label{sec:background}
\subsection{Statistical Language Models}
Language modeling fundamentally involves the construction of probability distributions over sequences of words or tokens. 
Typically, the conditional probability $P(t_i|t_1, ..., t_{i-1})$ denotes the likelihood that the token $t_i$ will follow the sequence of preceding tokens $h=t_1, ..., t_{i-1}$. 
However, accounting for entire sequences as context can become computationally prohibitive due to the possible vastness of sequence variation and length.
Hence, the n-gram model simplifies it by considering only the most recent $n-1$ tokens, adhering to a Markovian assumption:
\begin{equation}
    P_{\textit{ngram}}(t_i|h) = P(t_i|t_{i-n+1}, ..., t_{i-1})
\end{equation}
The n-gram model has been effectively used to capture highly repetitive regularities in both natural and programming languages, contributing significantly to various code-related tasks such as code completion~\cite{tu2014localness}, bug detection~\cite{jit}, and naturalness measure~\cite{buggy_code_naturalness}.
Subsequently, many pre-trained language models 
(e.g., CodeBERT~\cite{feng2020codebert}) 
are proposed, expanding on the capabilities of language modeling by accepting fixed-length token sequences. 
They leverage vast data and provide a user-friendly, out-of-the-box solution. 
Their robust learning mechanisms enable them to capture intricate code patterns, which are beneficial for a variety of tasks.

\subsection{Code Naturalness}
\label{sec:existing-methods}
As introduced in Section ~\ref{sec:introduction}, code naturalness is an important property of source code. 
Therefore, studying the accurate measure of code naturalness is critical. 
Hindle et al.~\cite{software_naturalness} were the first to measure code naturalness based on the n-gram model, calculating the cross-entropy for a line of code $S = t_1t_2...t_N$ as follows:
\begin{equation}
H_{\textit{ngram}}(S) = -\frac{1}{N}\sum_{i=1}^{N} \log P_{\textit{ngram}}(t_i|h)
\end{equation}
\revise{Here, $h$ represents the token sequence before $t_i$ in $S$ (i.e., $h=t_1, ..., t_{i-1}$).}

Then, Tu et al.~\cite{tu2014localness} improved the code naturalness measure by exploiting code localness and proposed the cached n-gram model:
\begin{equation}
\label{for:cache}
P_{\textit{cache}}(t_i|h) = \lambda \cdot P_{\textit{global}}(t_i|h) + (1 - \lambda) \cdot P_{\textit{local}}(t_i|h)
\end{equation}
In Formula ~\ref{for:cache}, $P_{\textit{global}}$ represents the conditional probability computed by the traditional n-gram model, which is trained with a large-scale corpus of source code, such as one project, whereas $P_{\textit{local}}$ represents the \textit{n-gram} model that is trained over small-scale source code, such as the context of a single line in a file.
$\lambda$ is the weight reflecting both the commonalities among source code in the large-scale corpus and the particularity under certain contexts.
The subsequent code naturalness calculation shares the same workflow as the one proposed by Hindle et al.~\cite{software_naturalness}.
Later, Ray et al.~\cite{buggy_code_naturalness} optimized its line naturalness measure by normalizing the computation for different types of code lines.

To overcome the limitation of n-gram models (e.g., trained on small datasets and thus inapplicable to cross-project scenarios), Khanfir et al.~\cite{khanfir2022codebert-nt} proposed \codebert{} by using the pre-trained CodeBERT model to measure code naturalness based on token predictability. 
\codebert{} masks some tokens in each individual line and uses the minimal prediction confidence of CodeBERT on the masked tokens as the naturalness of this line.

To sum up, all these methods measure code naturalness by applying the used statistical language model to measure the naturalness of each line individually.
However, they ignore essential code dependency information and thus may lead to inaccurate measure of code naturalness, since the dependency information is crucial for code comprehension as discussed in Section~\ref{sec:introduction}.

In what follows, we will use a simple example to illustrate this issue.
Considering the code snippet shown in Figure ~\ref{fig:overview} (simplified for clarity), 
the code is regarded as natural when each line of code is measured individually by existing methods since each code line is common and widely-used in the code snippet. 
However, after putting all the code lines together, it is indeed unnatural because the object {\tt example} at line 3 will be de-referenced when it is {\tt null}. 
Considering all the lines sequentially may not capture such information well since many lines between lines 1 and 2 (also after line 4) may involve too much noise and thus weaken the effect of code dependency. 
In such a case, incorporating accurate code dependency for measuring code naturalness may be helpful. 
In this work, we aim to conduct the first empirical exploration to investigate the influence of code dependency on measuring code naturalness by designing an initial (but reasonable) solution.

\section{Methodology}
\label{sec:approach}

\begin{figure*}[t]
    \centering
    \includegraphics[width=0.95\linewidth]{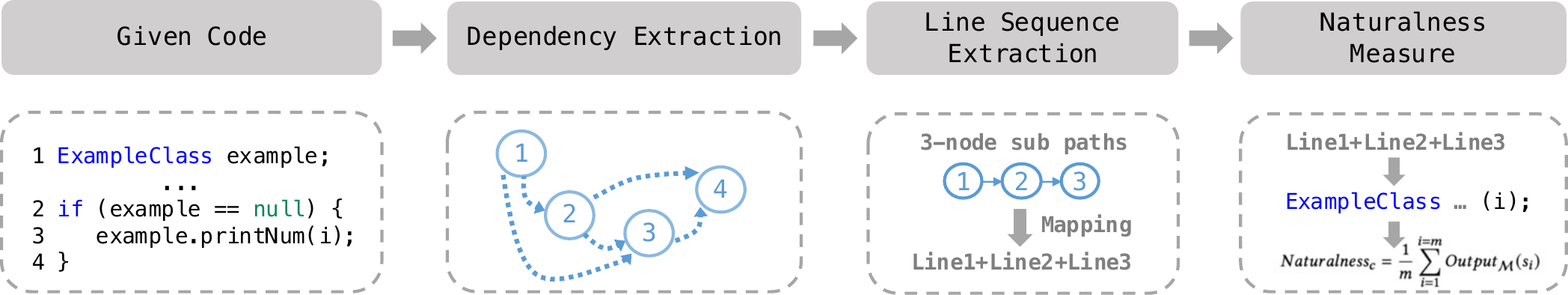}
\caption{Overview of \tech{}}
\label{fig:overview}
\vspace{-0.2cm}
\end{figure*}

In this work, we conduct the first empirical study to investigate the influence of code dependency on code naturalness.
To complete this study, we propose the first method of incorporating code dependency into measuring code naturalness, i.e., dependency-aware code naturalness (\tech{}).
In fact, there may be some other methods of utilizing code dependency, but our main contribution lies in the first exploration in this direction even though the idea of \tech{} is simple.
If we can obtain positive conclusions from the study with \tech{}, designing more advanced methods of incorporating code dependency can be promising future work.
In this section, we introduce \tech{} in detail.

In general, \tech{} consists of three steps as shown in Figure~\ref{fig:overview}.
(1) It represents the given code as a program dependency graph (PDG) by performing control- and data- flow analysis, which can help understand the semantics of code.
(2) It traverses all paths in the PDG, each of which can be treated as a sequence of lines with control or data dependency. It can incorporate code dependency without much noise compared to treating all lines in code as a sequence for measuring code naturalness.
(3) It applies the widely-used statistical model (i.e., n-gram or CodeBERT) to each sequence, instead of each individual line, for measuring code naturalness.
With \tech{}, code naturalness can be measured by incorporating code semantics with the aid of code dependency information, which can intuitively improve the precision of measuring code naturalness.
Next, we present dependency extraction in Section~\ref{sec:graph-extract}, line sequence extraction in Section~\ref{sec:code-grouping}, and naturalness measure in Section~\ref{sec:calculation}.

\subsection{Dependency Extraction} 
\label{sec:graph-extract}

\begin{figure}[t]
  \centering
  \includegraphics[width=0.6\linewidth]{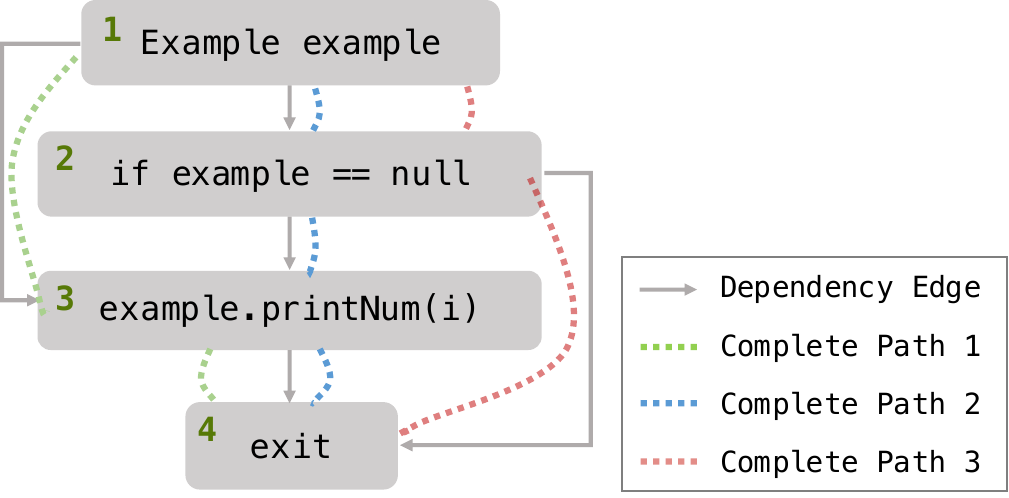}
  \caption{An example of program dependency graph}
  \label{fig:graph-example}
\end{figure}

To extract the dependency information for a given code snippet, \tech{} transforms the code into a PDG. 
The PDG serves as a representation of a program's control and data flow dependencies among program elements.
The PDG is represented as a directed graph $\mathcal{G} = (\mathcal{V}, \mathcal{E})$, where:
\begin{itemize}
    \item $\mathcal{V}$ is the set of nodes in the graph, and each node corresponds to a program element, e.g., statement or function.
    \item $\mathcal{E}$ is the set of directed edges between nodes in $\mathcal{V}$, representing the dependencies between program elements. 
\end{itemize}
Specifically, \tech{} constructs the PDG at the granularity of program statements. That is, each node $v$ in the graph corresponds to a statement in the source code, and the edges in the graph depict the dependencies between these statements.
For instance, considering the code displayed in Figure~\ref{fig:overview}, the PDG constructed at the statement granularity is shown in Figure~\ref{fig:graph-example}.
For \tech{}, there are two types of edges in the PDG:
\begin{itemize}
    \item $\mathcal{E}_c$: This type of dependency represents control flow relationship between statements. For example, if a statement A depends on another statement B in a way that the execution of A depends on the outcome of B, then a control dependency edge is present from B to A in the PDG. For example, the edge between node $v_2$ and node $v_3$ in Figure~\ref{fig:graph-example} represents the control dependency between them.
    \item $\mathcal{E}_d$: This type of dependency represents data flow relationship between statements. If a variable's value is used in another statement, a data dependency edge is established from the statement using the variable to the one defining its value. For example, the edge between node $v_1$ and node $v_3$ represents the data dependency of the object {\tt example}.

\end{itemize}

During the PDG construction process, \tech{} also analyzes the mapping between the nodes in the graph and the lines in source code, serving as the foundation for subsequent naturalness measure. 
For instance, node $v_1$ in the PDG is mapped back to the object declaration at Line 1 of the source code in Figure~\ref{fig:overview}.
However, it is essential to note that not all nodes in the PDG have corresponding source lines. For example, node $v_4$ in the PDG does not correspond to any lines in the source code. 
Also, multiple nodes in the graph sometimes correspond to the same line, such as a line of a {\tt for} statement that is represented by multiple nodes in the PDG.

\subsection{Line Sequence Extraction}
\label{sec:code-grouping}
Each path in the PDG represents a sequence of nodes that have dependencies.
To incorporate dependency information for measuring code naturalness, \tech{} extracts paths from the PDG. The extraction process is performed using a Depth-First Search (DFS) strategy, which starts from a designated starting node and explores each path throughout the graph systematically. 
When \tech{} reaches a node with no successors or the node representing the termination of code execution (i.e., the exit node), this path represents a complete path.
To handle cyclic paths, \tech{} maintains a set of visited nodes, ensuring that previously visited nodes are not revisited during the traversal process. 
In this way, \tech{} obtains a list of complete paths. 
For instance, \tech{} extracts three complete paths from the PDG of the code in Figure~\ref{fig:overview}, each of which is highlighted in different colors as shown in Figure~\ref{fig:graph-example}.
Note that \tech{} filters out the paths that are subsumed by any other paths.

However, complete paths are typically lengthy and intricate, and thus there are usually a lot of overlapped nodes between complete paths.
That results in redundant use of the dependencies between these nodes, which could \revise{incur extra computation overhead} and affect the precision of code naturalness measure.
Inspired by the existing work~\cite{rahman2019natural_revisited} that highlighted the repetition of n-node sub-graphs of source code, \tech{} addresses this problem by extracting n-node sub-paths from each complete path. 
For example, given n=3, four distinct sub-paths can be extracted from the three complete paths in the graph shown in Figure~\ref{fig:graph-example}, i.e., $\{v_1-v_2-v_3,$ $v_2-v_3-v_4,$ $v_1-v_3-v_4,$ $v_1-v_2-v_4\}$. 
Each sub-path represents a shorter sequence of nodes with dependencies.

To address the difficulty of using statistical models to measure the naturalness of PDG nodes,
\tech{} further leverages the mapping obtained in the first step to convert each sub-path into a sequence of corresponding code lines. 
Similarly, if a sequence is a subset or duplicate of any other sequences, \tech{} filters it out. 
For instance, if there are three extracted sequences: Line1-Line2-Line3, Line2-Line3, and Line2-Line3, \tech{} retains only the sequence Line1-Line2-Line3.
Since multiple nodes may correspond to the same code line, there may be the cases where the same line continuously appears in a sequence (e.g., Line3-Line5-Line5-Line5).
Actually, this line has encompassed the dependencies of all its corresponding nodes, and thus \tech{} retains only one occurrence of this line in the sequence to avoid redundancy (e.g., the above example sequence is reduced as Line3-Line5).

This process results in a refined set of sequences, each of which consists of code lines with data or control dependencies.
In this way, \tech{} can incorporate code dependency into naturalness measure by applying the widely-used statistical model to each sequence of code lines.

\subsection{Naturalness Measure}
\label{sec:calculation}
After obtaining the set of sequences of code lines, \tech{} follows the similar workflow of the existing methods~\cite{buggy_code_naturalness, khanfir2022codebert-nt} to measure the naturalness of code. 
To make our method description self-contained, we also introduce the workflow briefly in this section.
When measuring the naturalness of a given code snippet, \tech{} treats each sequence of code lines with dependencies as a unit for naturalness measure.
That is, \tech{} applies a statistical model to each sequence of code lines instead of each code line individually, and then aggregates the naturalness of all sequences to obtain the naturalness of the given code snippet.

Specifically, given that \tech{} reorganizes the given code snippet into m sequences of code lines: $c = \{s_1, s_2, \ldots, s_m\}$, where $s_i$ represents a sequence of code lines with data or control dependencies. 
Then, \tech{} measures the naturalness of the code as follows:
\begin{equation}
\textit{Naturalness}_c = \frac{1}{m} \sum_{i=1}^{i=m} \textit{Output}_{\mathcal{M}}(s_i)
\end{equation}
Here, $\mathcal{M}$ represents the statistical model used to measure code naturalness and $\textit{Output}_{\mathcal{M}}(s_i)$ represents the naturalness of $s_i$ measured by $\mathcal{M}$.
Note that the idea of \tech{} is general and can be applied independently of the used statistical model.
Therefore, we can use any statistical model in this step of \tech{}.
To demonstrate the generality of our \tech{} idea, we create two instantiations of \tech{} by integrating the two widely-used statistical models in measuring code naturalness (i.e., the cached n-gram model and the pre-trained CodeBERT model) for evaluation, respectively.
As described in Section~\ref{sec:introduction}, we call them \tech{}$_{\textit{ngram}}$ and \tech{}$_{\textit{codebert}}$, respectively.
The details of using the cached n-gram and pre-trained CodeBERT models to measure code naturalness have been introduced in Section~\ref{sec:background}.
As it is infeasible to determine one syntax type for a sequence of multiple code lines, \tech{}$_{\textit{ngram}}$ does not perform syntax-based normalization like the state-of-the-art n-gram-based method~\cite{buggy_code_naturalness}.

\section{Evaluation}
\label{sec:evaluation}
With \tech{}, we conducted the first empirical study to investigate whether incorporating code dependency can improve the measure of code naturalness.
We evaluated \tech{} (i.e., its two instantiations \tech{}$_{\textit{ngram}}$ and \tech{}$_{\textit{codebert}}$) in three scenarios for code naturalness, 
including its core task (i.e., distinguishing natural and unnatural code) - \textbf{RQ1}, 
and two applications of code naturalness:
distinguishing buggy and non-buggy code lines (called buggy line prioritization) - \textbf{RQ2}, and cleansing data for training code generation models (called training data cleansing) - \textbf{RQ3}. 
\del{
Finally, we also investigated the contribution of our code dependency extraction strategy in \tech{} by designing proper variants for comparison - \textbf{RQ4}.
}

\smallskip
\noindent$\bullet$
\textbf{Compared Techniques}.
\label{sec:compared_tech}
We chose the \textbf{\old{}}~\cite{buggy_code_naturalness} and \textbf{\codebert{}}~\cite{khanfir2022codebert-nt} methods for comparison in our study 
, which have been introduced in Section~\ref{sec:background}.
To investigate the contribution of incorporating code dependency to measure code naturalness clearly, we compared \tech{}$_{\textit{ngram}}$/\tech{}$_{\textit{codebert}}$ with \old{}/\codebert{} respectively, by controlling the influence of the used statistical model.
In particular, it is helpful to demonstrate the generality of our \tech{} idea across different statistical models.

As mentioned in Section~\ref{sec:introduction}, considering all code lines together in a sequential manner is an intuitive method to include all dependency information within the given code, but can incur much noise in code dependency. 
We regarded it as a compared technique in the study, which can help investigate the contribution of our code dependency extraction strategy in \tech{}. 
For ease of presentation, we call it \textbf{\oldvariant{}} (short for complete sequence code naturalness). 
Due to the input length limitation in CodeBERT, we cannot use CodeBERT as the statistical model in \oldvariant{}, and thus we just constructed the instantiation of \oldvariant{} by using the cached n-gram model as the statistical model (called \textbf{\oldvariant{}$_{\textit{ngram}}$}).

Besides, \tech{} extracts n-node sub-paths from each complete path to relieve the lengthiness and intricacy of complete paths.
\oldvariant{} may also suffer from this issue.
Hence, we improve \oldvariant{} by splitting the complete sequence with all code lines used in it into a set of n-line subsequences, then measuring naturalness of each n-line subsequence, and finally aggregating the naturalness of all subsequences, for more sufficient comparison.
We call this variant of \textbf{\oldvariant{}-sub}.
According to the used statistical model, we created two instantiations, i.e., \textbf{\oldvariant{}-sub$_{\textit{ngram}}$} and \textbf{\oldvariant{}-sub$_{\textit{codebert}}$}.
\revise{
We set n to 3 by default following the existing study~\cite{rahman2019natural_revisited}, which demonstrated the repetitiveness of 3-node sub-graphs of code, and extracted all sub-paths with length 3 from complete paths. 
However, we also investigated the influence of n by setting n to 2 in \tech{}.
}

\revise{
Overall, we had two sets of comparisons: \tech{}$_{\textit{ngram}}$ vs. \tech{}$_{\textit{ngram}}$(n=2)  vs. \old{} v.s. \oldvariant{}$_{\textit{ngram}}$ v.s. \oldvariant{}-sub$_{\textit{ngram}}$ and \tech{}$_{\textit{codebert}}$ vs. \tech{}$_{\textit{codebert}}$(n=2) vs. \codebert{} v.s. \oldvariant{}-sub$_{\textit{codebert}}$, so as to evaluate the importance of incorporating code dependency in measuring code naturalness regardless of the used statistical models.
As fine-tuning code models is resource-intensive and time-consuming, we only compared \tech{}$_{\textit{ngram}}$ vs.  \old{} and \tech{}$_{\textit{codebert}}$ vs. \codebert{} for the third scenario (i.e., training data cleansing).
}

\smallskip
\noindent$\bullet$
\textbf{Implementations}.
We implemented \tech{} in Java and Python.
We implemented PDG construction using Soot~\cite{vallee1999soot} (a widely-used program analysis tool), and constructed the mapping between nodes and code lines based on the debugging information provided by Soot (i.e., line number tables and local variable tables).
\del{
For n-node sub-path extraction in \tech{}, we set n to 3 by default following the existing study~\cite{rahman2019natural_revisited}, which demonstrated the repetitiveness of 3-node sub-graphs of code,
and extracted all sub-paths with length 3 from complete paths. 
For more fair comparison, we set n in \oldvariant{}-sub (the step of n-line subsequence extraction) to 3 as well.
}
We implemented the cached n-gram model based on nltk 3.8.1~\cite{loper2002nltk}, javalang 0.13.0~\cite{javalang}, and scikit-learn 1.2.2~\cite{sklearn}.
For \old{}, we replicated the experimental results in the original paper~\cite{buggy_revisited} based on our re-implementation, which validates the correctness of our re-implementation to a large extent.
For \codebert{}, we directly used their released tool~\cite{khanfir2022codebert-nt}.

Although the idea of \tech{} is general, we implemented and evaluated it on Java projects currently, which may not represent the subjects under other programming languages. 
This is a potential threat in our study, and we discussed it in detail in Section~\ref{sec:threats}.
Our decision to first focus on Java projects was influenced by the high-impact study conducted by Ray et al.~\cite{buggy_code_naturalness} and Khanfir et al. ~\cite{khanfir2022codebert-nt}, which were also carried out on Java projects. 
This choice allows for more direct comparison between our study and the existing work. 

We conducted all the experiments on a server with Intel(R) Xeon(R) Silver 4214 @ 2.20GHz CPU, 256GB memory, and Ubuntu 18.04 operating system.

\subsection{RQ1: Effectiveness Comparison on Distinguishing Natural and Unnatural Code}
\label{sec:rq1}
\subsubsection{Dataset}
To answer RQ1, it is essential to create a collection of both natural code and unnatural code.
Here, we treated the code in the HumanEval-X dataset~\cite{zheng2023codegeex}, which extends the widely-used HumanEval dataset~\cite{chen2021evaluating} to multiple languages (including the Java version we focused on), as natural code.
This dataset consists of 164 hand-written programming problems and the corresponding answers, which has been widely used as the reference to evaluate code quality~\cite{athiwaratkun2022multi,zheng2023codegeex,nijkamp2022codegen,touvron2023llama}.
\revise{In HumanEval-X, 34.99\% of statements are dependent on other statements and each dependency sequence involves 7 statements on average.}
The code in HumanEval-X is meticulously crafted by experienced developers and has been recognized as high-quality code~\cite{athiwaratkun2022multi, zheng2023codegeex}.
Hence, this dataset is suitable as natural code in this experiment.

We then constructed the corresponding unnatural code from each piece of natural code in HumanEval-X. 
Specifically, we implemented a set of semantic-preserving transformation rules for de-naturalizing code following the existing work~\cite{natgen, mutate_motivation2}.
By applying them to a piece of natural code, we can obtain a set of unnatural but equivalent code. 
The reasons for obtaining unnatural code in this way are two-fold:
(1) There is no existing open-source dataset of unnatural code to our best knowledge;
(2) Such a way can control the difference between each pair of natural code and unnatural code, which can help evaluate the ability of distinguishing natural code and unnatural code more purely by keeping the remaining part of code the same (except the de-naturalized part of code).

In total, we implemented three equivalent transformation rules to de-naturalize code. 
The general idea of these rules is to transform frequently-encountered natural code that developers can typically write, into the synthetic unnatural form. 
The existing studies~\cite{casalnuovo2020programmers,casalnuovo2020does} have demonstrated that the de-naturalized code with these transformation rules is more challenging to read and understand.
We used an example shown in Figure~\ref{fig:mutation} to facilitate illustrating the three rules in the following:

\begin{figure*}[t]   
    \centering     
    \begin{subfigure}[t]{0.35\textwidth}
    \centering
        \includegraphics[width=\linewidth]{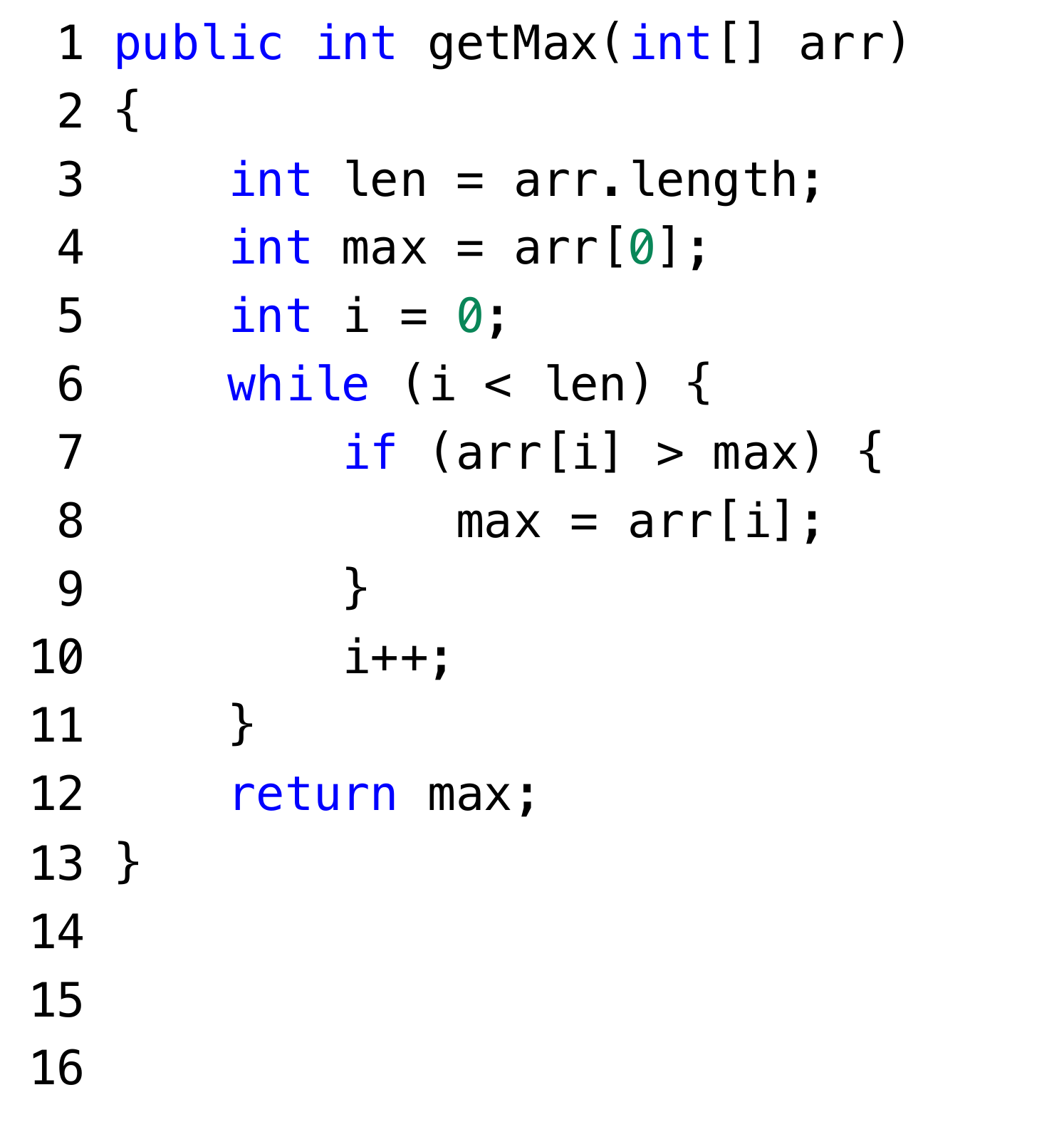}
        \caption{Original code}
        \label{fig:origin}
    \end{subfigure}
     \hspace{5mm} 
    \begin{subfigure}[t]{0.351\textwidth}
    \centering
        \includegraphics[width=\linewidth]{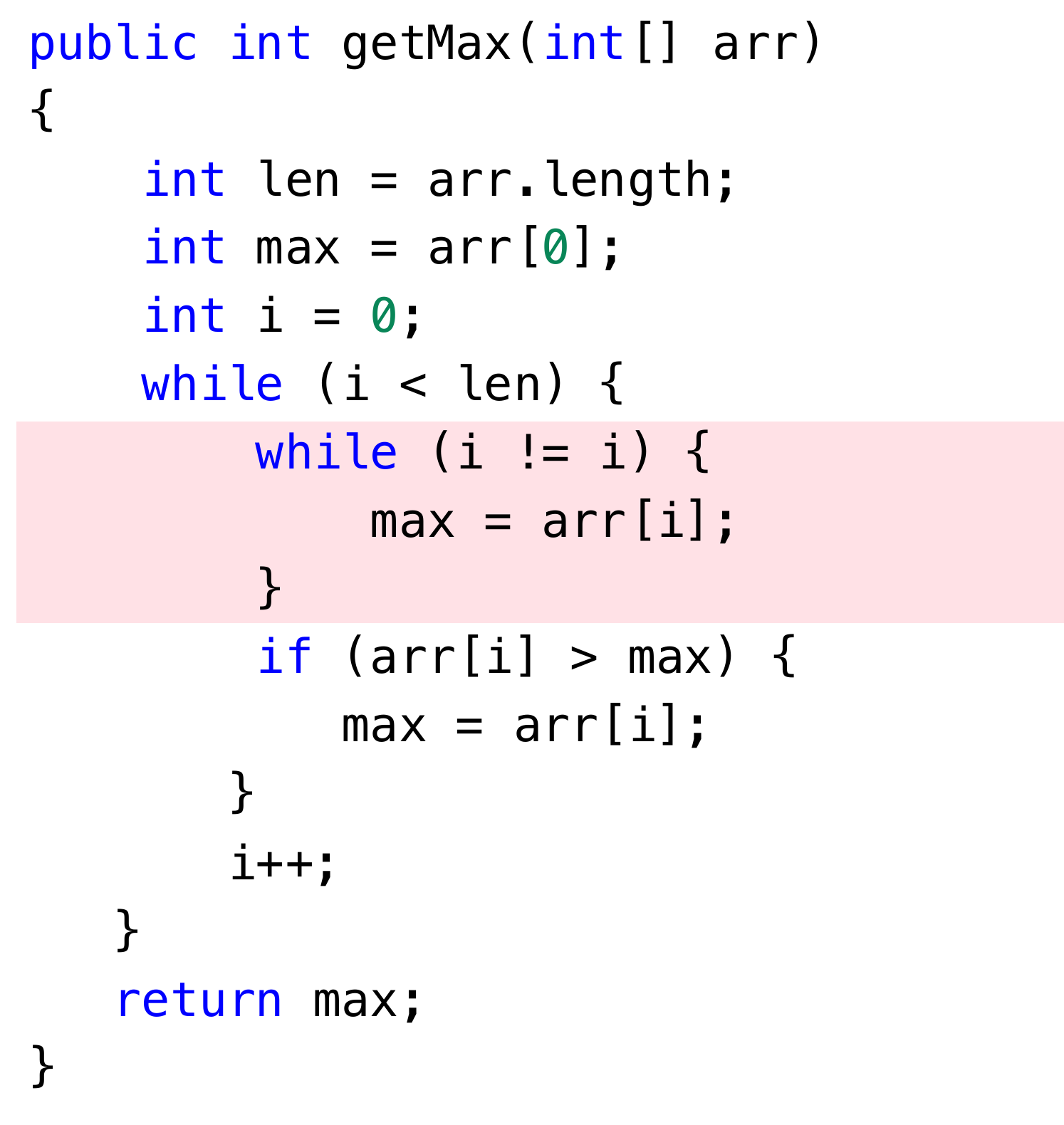}
        \caption{Dead Code Insertion}
        \label{fig:deadcode}
    \end{subfigure}
    \\
    \vspace{2mm}
    \begin{subfigure}[t]{0.35\textwidth}
    \centering
        \includegraphics[width=\linewidth]{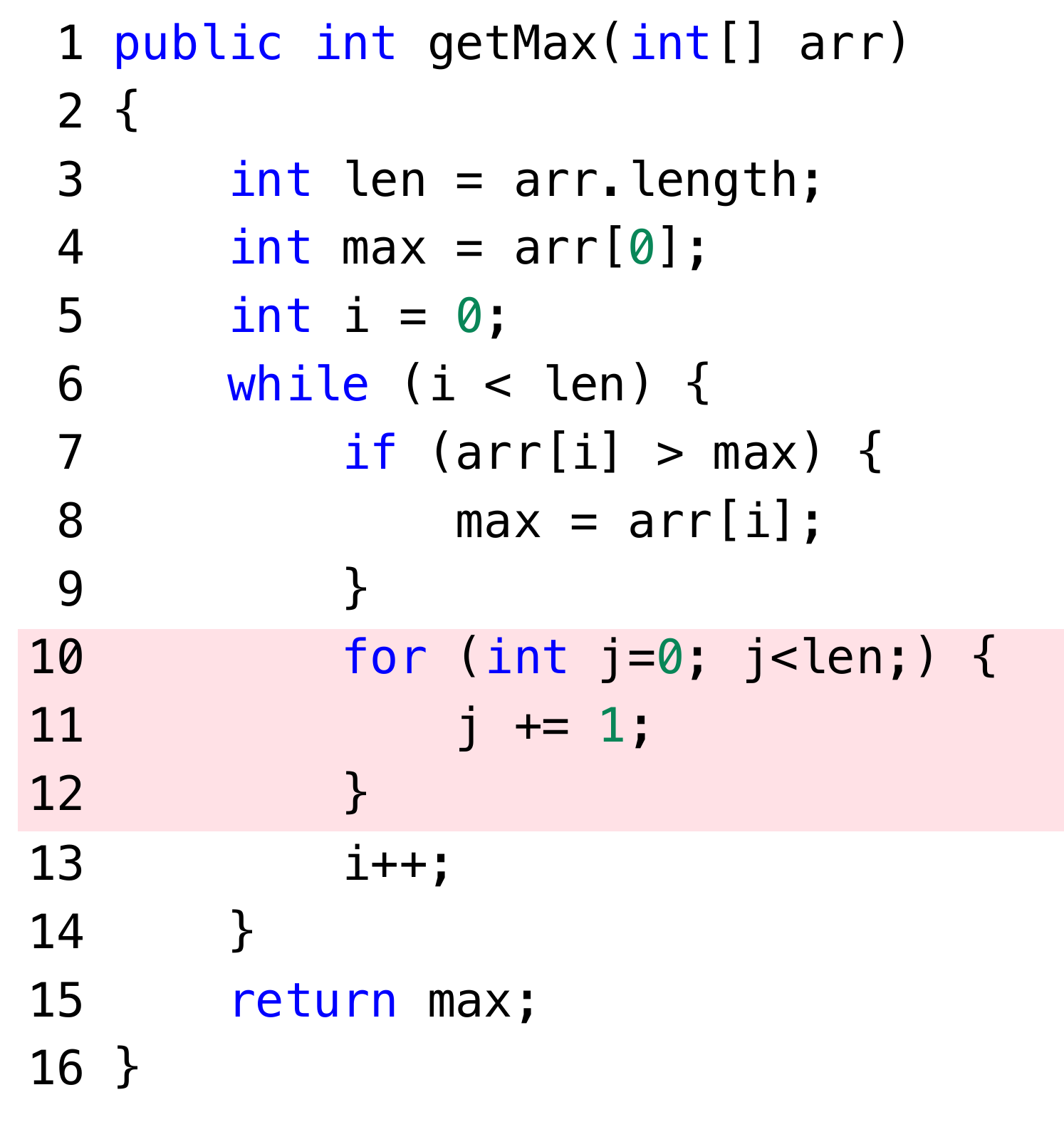}
        \caption{Confusing Code Insertion}
         \label{fig:confusing}
    \end{subfigure}
     \hspace{5mm} 
    \begin{subfigure}[t]{0.355\textwidth}
    \centering
        \includegraphics[width=\linewidth]{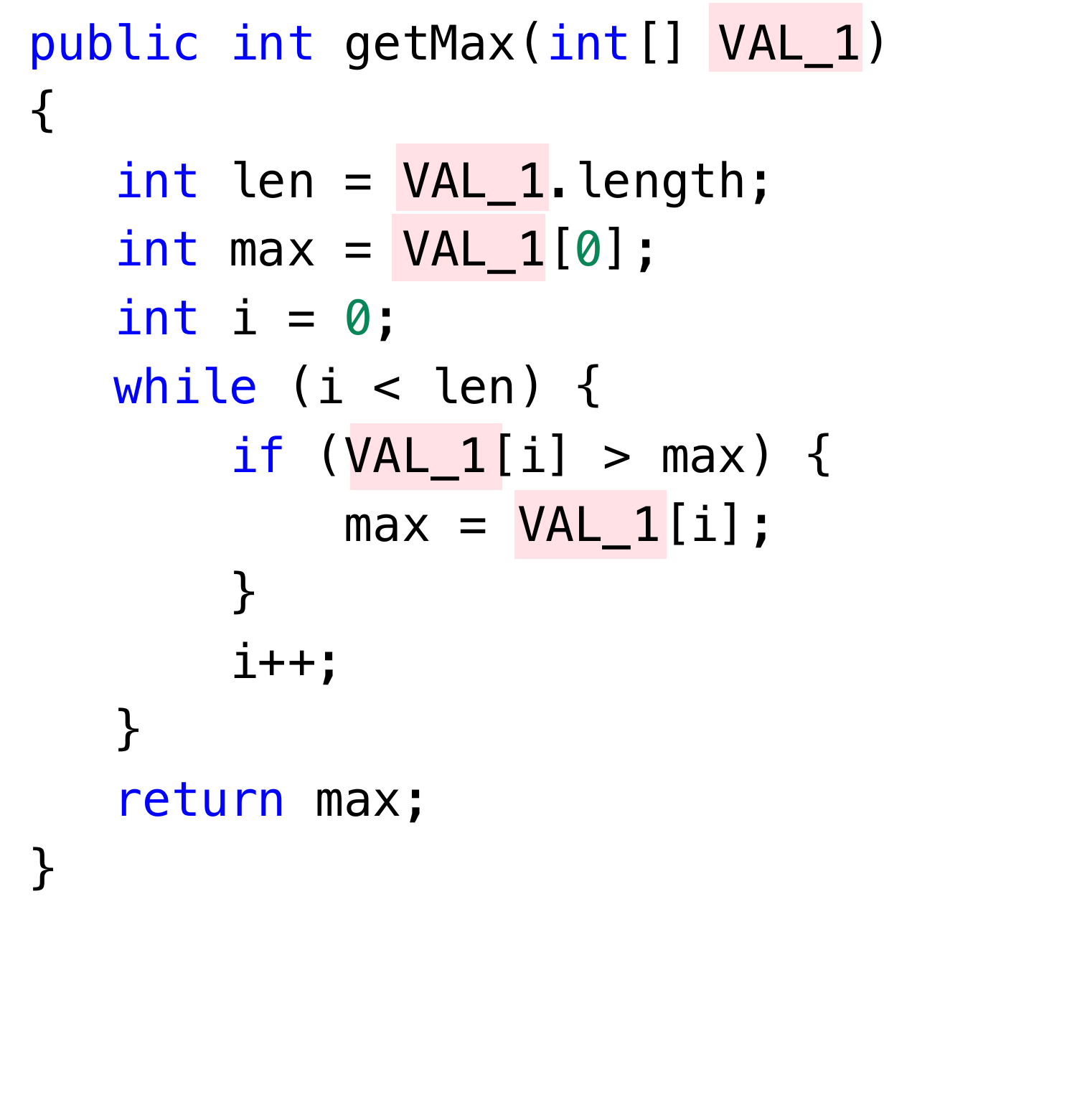}
        \caption{Variable Renaming}
        \label{fig:rename}
    \end{subfigure}
    \caption{Examples of de-naturalization}
    \label{fig:mutation}
\end{figure*}

\textbf{Dead Code Insertion}:
It inserts a block of dead code that will never be executed at random positions in the original code. 
To ensure the property of dead code, the inserted block is a loop or branch block, which is guarded by an unsatisfiable condition (e.g., a variable is not equal to itself).
The content in the branch or loop body is the statements transplanted from the original code. 
As shown in Figure~\ref{fig:deadcode}, a {\tt while} loop with an identically-false condition is inserted at Lines 7-9.

\textbf{Confusing Code Insertion}:
It inserts a block of code, which can be executed without altering the original functionality, at random positions in the natural code.
Specifically, the inserted block is a loop that can be executed several times and the content in the loop body is a statement with increment or decrement operation on a variable.
For example, it inserts a {\tt for} loop that can be executed without altering the original functionality at Lines 10-12 as shown in Figure~\ref{fig:confusing}.

\textbf{Variable Renaming}:
It renames variables to {\tt VAR\_i} (an unnatural name). 
When renaming a variable, we analyze the data-flow of this variable and rename all the occurrences of this variable in the original code. 
For example, we rename the variable {\tt arr} in Figure~\ref{fig:origin} to {\tt VAL\_1} in Figure~\ref{fig:rename}.

For each piece of natural code in HumanEval-X, we applied a transformation rule to create four distinct unnatural versions, each exhibiting different degrees of de-naturalization. 
They are produced based on $k$-order (k=\{1,2,3,4\}) transformation with a specific transformation rule (applying the transformation rule $k$ times to the piece of natural code), respectively.
Consequently, we crafted four sets of unnatural code per transformation rule from HumanEval-X, accumulating 12 sets across three distinct rules. 
This comprehensive collection of data is instrumental in assessing how well different methods can distinguish natural code and its unnatural counterparts with different degrees of de-naturalization.
\revise{
Note that we did not use loop transformation and block/operation-swap transformation in the existing work~\cite{natgen}, since they cannot guarantee the transformed code is really unnatural (e.g., changing a==b to b==a). Regarding these two rules, code naturalness and readability can be retained to a large extent since they retain structural information to a large extent and all textual semantics without changing identifiers.
}

\subsubsection{Metrics}
To evaluate the ability of distinguishing natural code and unnatural code for each method of measuring code naturalness, we first measured the naturalness of each piece of natural code and the corresponding unnatural code transformed from it in each set of unnatural code.
For ease of presentation, we denote the naturalness of the original natural code as $\textit{Naturalness}_n$ and denote that of the corresponding unnatural one as $\textit{Naturalness}_u$.
Then, we measured the normalized difference in naturalness between each pair of natural and unnatural code: $\textit{diff}$ = $\frac{\textit{Naturalness}_u - \textit{Naturalness}_n}{\textit{Naturalness}_n}$.
\revise{The difference provides insights into how distinctly a method can capture changes in code naturalness, reflecting its sensitivity to a large extent. 
Also, larger differences provide higher confidence to trust measure results to some degree. If a piece of code becomes unnatural after (slight) code changes during evolution, the measure with higher sensitivity is more effective to identify it. }

\subsubsection{Results and Analysis}
\begin{table*}[t]
\renewcommand\arraystretch{1.0} 
\caption{Effectiveness on distinguishing natural and unnatural code}
\label{tab:rq2-mutation-1}
\resizebox{0.95\linewidth}{!}{
\renewcommand\arraystretch{1.1} 
\begin{threeparttable}
\begin{tabular}{l|rrrr|rrrr|rrrr}
\toprule
\multicolumn{1}{c|}{\multirow{2}{*}{\textbf{Method}}} &
  \multicolumn{4}{c|}{\textbf{Dead Code Insertion}} &
  \multicolumn{4}{c|}{\textbf{Confusing Code Insertion}} &
  \multicolumn{4}{c}{\textbf{Variable Renaming}}
  \\
  \cmidrule(l){2-13} 
\multicolumn{1}{c|}{} &
  \multicolumn{1}{c}{\textbf{1}} &
  \multicolumn{1}{c}{\textbf{2}} &
  \multicolumn{1}{c}{\textbf{3}} &
  \multicolumn{1}{c|}{\textbf{4}} &
  \multicolumn{1}{c}{\textbf{1}} &
  \multicolumn{1}{c}{\textbf{2}} &
  \multicolumn{1}{c}{\textbf{3}} &
  \multicolumn{1}{c|}{\textbf{4}} &
  \multicolumn{1}{c}{\textbf{1}} &
  \multicolumn{1}{c}{\textbf{2}} &
  \multicolumn{1}{c}{\textbf{3}} &
  \multicolumn{1}{c}{\textbf{4}} 
  \\
  \midrule
  \tech{}$_{\textit{ngram}}$ &
  \textbf{12.17\%} &
  \textbf{19.08\%} &
  \textbf{23.53\%} &
  \textbf{26.26\%} &
  \textbf{24.86\%} &
  \textbf{38.76\%} &
  \textbf{46.88\%} &
  53.24\% &
  \textbf{17.70\%} &
  30.12\% &
  \textbf{42.23\%} &
  \textbf{48.45\%} 
  \\
  \tech{}$_{\textit{ngram}}$(n=2) &
  11.30\% &	
  18.01\%	&
  22.50\%	&
  25.29\%	&
  24.68\% &	
  38.11\% &
  46.31\% &	
  \textbf{53.72\%}	&
  17.65\%	&
  \textbf{30.21\%}	&
  42.16\%	& 
  48.05\% \\
  \old{} &
  7.96\% &
  14.05\% &
  19.95\% &
  22.82\% &
  17.96\% &
  29.46\% &
  37.60\% &
  42.97\% &
  9.13\% &
  19.09\% &
  25.12\% &
  31.72\% 
\\
\oldvariant{}$_{\textit{ngram}}$ &
  8.58\% &
  10.66\% &
  15.39\% &
  15.80\% &
  19.88\% &
  31.05\% &
  37.85\% &
  43.40\% &
  14.18\% &
  20.29\% &
  32.85\% &
  35.96\% 
  \\
\oldvariant{}-sub$_{\textit{ngram}}$ &
  10.06\% &
  16.66\% &
  20.39\% &
  25.33\% &
  17.44\% &
  25.87\% &
  34.16\% &
  38.78\% &
  6.77\% &
  14.30\% &
  18.78\% &
  26.13\% 
  \\
  \midrule
\tech{}$_{\textit{codebert}}$ &
  \textbf{9.46\%} &
  10.11\% &
  \textbf{13.20\%} &
  13.75\% &
  \textbf{19.37\%} &
  \textbf{30.41\%} &
  \textbf{32.43\%} &
  \textbf{32.68\%} &
  \textbf{14.83\%} &
  \textbf{26.28\%} &
  \textbf{35.65\%} &
  \textbf{37.87\%} 
   \\
\tech{}$_{\textit{codebert}}$ (n=2) &
  8.90\% &
  9.62\% &
  11.15\% &
  12.87\% &
  8.48\% &
  15.51\% &
  18.28\% &
  16.05\% &
  10.72\% &
  17.56\% &
  32.75\% &
  36.85\% 
  \\
\codebert{} &
  7.90\% &
  5.25\% &
  7.61\% &
  7.06\% &
  2.02\% &
  -0.25\% &
  0.35\% &
  0.74\% &
  10.59\% &
  15.70\% &
  29.49\% &
  35.91\% 
  \\
\oldvariant{}-sub$_{\textit{codebert}}$ &
  9.35\% &
  \textbf{13.50\%} &
  10.79\% &
  \textbf{14.66\%} &
  10.33\% &
  8.26\% &
  6.59\% &
  5.91\% &
  10.28\% &
  18.43\% &
  28.37\% &
  28.96\% 
  \\
\bottomrule
\end{tabular}
\end{threeparttable}
}
\end{table*}

Table~\ref{tab:rq2-mutation-1} shows the comparison effectiveness of \tech{}$_{\textit{ngram}}$ v.s. \old{} and \tech{}$_{\textit{codebert}}$ v.s. \codebert{} in distinguishing natural code and unnatural code produced by different orders of transformation with different rules.
For each set of unnatural code, we presented the average normalized difference values and put all the detailed results at our project homepage.
From Table~\ref{tab:rq2-mutation-1}, we found that \tech{}$_{\textit{ngram}}$ outperforms \old{}, and \tech{}$_{\textit{codebert}}$ outperforms \codebert{} across different degrees of unnatural code produced by different transformation rules.
Specifically, \tech{}$_{\textit{ngram}}$ improves upon \old{} by an average of 41.82\%, while \tech{}$_{\textit{codebert}}$ improves upon \codebert{} by an average of 13.41 times, in terms of average normalized difference values across all the 12 sets of unnatural code constructed from the set of natural code.
In particular, \codebert{} even incorrectly identifies unnatural code as being more natural in one case (i.e., the average normalized difference value is negative), whereas \tech{}$_{\textit{codebert}}$ avoids such misjudgments. 
The results demonstrate the superiority of \tech{} in distinguishing natural and unnatural code regardless of the used statistical models.
In particular, we performed a paired sample Wilcoxon signed-rank test~\cite{woolson2007wilcoxon} at the significant level of 0.05 to further confirm the superiority of \tech{} over the state-of-the-art counterparts in statistics by obtaining the p-values all smaller than 0.05.
\revise{
The effect size between \tech{}$_{\textit{ngram}}$/\tech{}$_{\textit{codebert}}$ and Ngram-NT/CodeBERT-NT is 0.36 and 0.65. 
}

\revise{
Moreover, the results also demonstrate that \tech{}$_{\textit{ngram}}$ outperforms \oldvariant{}-sub$_{\textit{ngram}}$ and \tech{}$_{\textit{codebert}}$ outperforms \oldvariant{}-sub$_{\textit{codebert}}$ in the vast majority of cases.
On average, \tech{}$_{\textit{ngram}}$ shows an improvement of 58.65\% over \oldvariant{}-sub$_{\textit{ngram}}$, and \tech{}$_{\textit{codebert}}$ shows an average improvement of 136.75\% over \oldvariant{}-sub$_{\textit{codebert}}$,  in terms of average normalized difference values.
In additional, \tech{}$_{\textit{ngram}}$ always performs better than \oldvariant{}$_{\textit{ngram}}$.
For different n values, DAN with n=3 outperforms that with n=2 in the vast majority of cases, but the latter still significantly outperforms the baselines.
The results not only confirm the contribution of relatively broad dependency information compared to the local context information incorporated by \oldvariant{}-sub, but also confirm the poor performance of directly treating the code as a whole sequence due to incurring too much noise (even though including all dependency information), demonstrating the importance of incorporating code dependency in a reasonable manner for measuring code naturalness (like \tech{}).
}

Here, we noticed that CodeBERT-based methods (\tech{}$_{\textit{codebert}}$ and \codebert{}) perform worse than n-gram-based methods (\tech{}$_{\textit{ngram}}$ and \old{}) in many cases.
The reason may be that CodeBERT was pre-trained on extensive open-source projects, which may not work well on our datasets constructed with minimal de-naturalization, while the cached n-gram model considers localness well.

We also constructed four sets of unnatural code by mixing the three rules for $k$-order (k=\{1,2,3,4\}) transformation on each piece of natural code for more sufficient evaluation.
Specifically, we randomly selected a rule from the three for each transformation.
All the conclusions are consistent with the above ones.
Due to the space limit, we put the detailed results at our project homepage.

\begin{tcolorbox}
\textbf{Finding 1:} 
Incorporating code dependency for measuring code naturalness with \tech{} achieves a stronger ability of distinguishing natural and unnatural code than the state-of-the-art \old{} and \codebert{} methods that isolate each code line for measuring code naturalness.
\revise{The code dependency extraction strategy contribute to the overall effectiveness of \tech{} compared to the intuitive methods of treating code as a whole sequence.}
\end{tcolorbox}

\subsection{RQ2: Effectiveness Comparison on Buggy Line Prioritization}
As demonstrated by the existing studies~\cite{buggy_code_naturalness, buggy_revisited}, buggy code lines often exhibit worse naturalness than non-buggy lines.
Therefore, code naturalness has been used to identify buggy lines for early inspection in order to improve the efficiency of software development and maintenance~\cite{khanfir2022codebert-nt,jit}.
In this RQ, we investigated whether incorporating code dependency into measuring code naturalness with \tech{} can help improve the effectiveness of buggy line prioritization.

Note that we do not aim to compete with the state-of-the-art static bug finders and fault localization techniques, but use this downstream application to demonstrate the superiority of \tech{}$_{\textit{ngram}}$ over \old{} and \tech{}$_{\textit{codebert}}$ over \codebert{}.

\subsubsection{Datasets}
To answer RQ2, we collected all the datasets used for buggy line prioritization from the existing studies~\cite{buggy_revisited,khanfir2022codebert-nt},
i.e., Defects4J~\cite{just2014defects4j}, GrowingBugs~\cite{growing}, and SmartSHARK~\cite{Herbold2020smartshark}.
Note that we are the first to use all the three datasets for evaluating code naturalness measures in the scenario of buggy line prioritization.
Each dataset contains a collection of real-world bugs from different Java projects.
In total, we used 1,346 real-world bugs in this experiment, including 614 bugs from Defects4J, 324 bugs from GrowingBugs, and 368 bugs from SmartSHARK.
\revise{Across all three datasets, 27.22\% of statements are dependent on others and each dependency sequence involves 25 statements per file.}
We discarded the bugs with compilation errors since the underlying tool (i.e., Soot) extracts code dependency at the Jimple code.
The main reason lies in that many bugs from SmartSHARK are very old and thus it is difficult for them to restore the environment and dependency required for compilation.
Nevertheless, the number of bugs used in our study is the largest among all the existing studies on buggy lines prioritization~\cite{buggy_code_naturalness, buggy_revisited}.
Each bug is equipped with a buggy version and a bug-fixed version.
Following the existing studies~\cite{buggy_code_naturalness,buggy_revisited}, we identified the code lines in the buggy version that have either been deleted or modified in the bug-fixed version as \textbf{buggy lines}.
The remaining lines in the same files as buggy lines are identified as \textbf{non-buggy lines}.

\subsubsection{Process}
Following the existing work~\cite{buggy_revisited, jit}, we applied \old{} and \codebert{} to each line in the buggy files for each bug, respectively.
It measures the naturalness of each line, and then prioritizes all the lines in the buggy files as the descending order of their measured naturalness scores.
The lines with higher priorities are more likely to be buggy.
For \tech{}$_{\textit{ngram}}$ and \tech{}$_{\textit{codebert}}$, we applied them to measure the naturalness of each sequence that consists of multiple lines with dependencies and then prioritized all the sequences in the buggy files in the descending order, respectively. 
To break the tie for the lines within each sequence, we prioritized them by measuring the naturalness of each of these lines same as \old{}/\codebert{}.

\subsubsection{Metrics}
In this experiment, we adopted the widely-used metrics, i.e., \textbf{RAUC-k} (Raw Area Under the Curve) and \textbf{MRR} (Mean Reciprocal Rank), following the existing studies on prioritization tasks~\cite{chen2023exploring, wang2021prioritizing, yang2023silent}.
RAUC-k measures the effectiveness of a prioritization technique for the first $k$ results in a prioritized list~\cite{wang2021prioritizing}.
Specifically, it transforms the prioritization result produced by a technique to a plot, where the x-axis represents the number of prioritized lines and the y-axis represents the number of identified buggy lines.
Then, it calculates the ratio of the area under the curve for a technique to that for the perfect prioritization.
To perform comprehensive comparison, we considered $k$ to be 20\%, 40\%, 60\%, 80\%, 100\% of the number of prioritized lines, respectively. 
MRR measures the effectiveness of a  prioritization technique based on its ability to rapidly identify the first correct result~\cite{mrr}. 
Specifically, MRR treats each buggy file as a query and computes the reciprocal of the rank of the first correctly identified buggy line for each query, and then averages these reciprocal ranks across all queries to measure the prioritization effectiveness. 
Larger RAUC-k and MRR values mean better prioritization results, indicating better effectiveness of measuring code naturalness in the application of buggy line prioritization.

\subsubsection{Results and Analysis}

\begin{table*}[]
\setlength{\abovecaptionskip}{0.3cm}
\caption{Effectiveness on buggy line prioritization}
\label{tab:rq3-prioritization}
\resizebox{1.0\linewidth}{!}{
\begin{threeparttable}
\begin{tabular}{l|llllll|llllll|llllll}
\toprule
\multicolumn{1}{c|}{\multirow{2}{*}{\textbf{Method}}} &
  \multicolumn{6}{c|}{\textbf{Defects4j}} &
  \multicolumn{6}{c|}{\textbf{GrowingBugs}} &
  \multicolumn{6}{c}{\textbf{SmartSHARK}} \\ 
  \cmidrule{2-19} 
\multicolumn{1}{l|}{} &
  \multicolumn{1}{c}{\textbf{20\%}} &
  \multicolumn{1}{c}{\textbf{40\%}} &
  \multicolumn{1}{c}{\textbf{60\%}} &
  \multicolumn{1}{c}{\textbf{80\%}} &
  \multicolumn{1}{c}{\textbf{100\%}} &
  \multicolumn{1}{c|}{\textbf{MRR}} &
  \multicolumn{1}{c}{\textbf{20\%}} &
  \multicolumn{1}{c}{\textbf{40\%}} &
  \multicolumn{1}{c}{\textbf{60\%}} &
  \multicolumn{1}{c}{\textbf{80\%}} &
  \multicolumn{1}{c}{\textbf{100\%}} &
  \multicolumn{1}{c|}{\textbf{MRR}} &
  \multicolumn{1}{c}{\textbf{20\%}} &
  \multicolumn{1}{c}{\textbf{40\%}} &
  \multicolumn{1}{c}{\textbf{60\%}} &
  \multicolumn{1}{c}{\textbf{80\%}} &
  \multicolumn{1}{c}{\textbf{100\%}} &
  \multicolumn{1}{c}{\textbf{MRR}} \\
  \midrule
\tech{}$_{\textit{ngram}}$ &
  \textbf{0.133} &
  \textbf{0.240} &
  \textbf{0.333} &
  \textbf{0.406} &
  \textbf{0.474} &
  \textbf{0.118} &
  \textbf{0.155} &
  \textbf{0.259} &
  \textbf{0.344} &
  \textbf{0.425} &
  \textbf{0.499} &
  \textbf{0.178} &
  \textbf{0.194} &
  \textbf{0.356} &
  \textbf{0.491} &
  \textbf{0.592} &
  \textbf{0.667} &
  \textbf{0.129} 
  \\
  \tech{}$_{\textit{ngram}}$(n=2) & 0.132 &	0.239 &	0.333	& 0.406 &	0.474 &	0.117 &	0.152	& 0.256	&  0.343	& 0.424	& 0.498	& 0.177& 	0.181&	0.343	&0.483&	0.585&	0.662&	0.126 \\
  \old{} &
  0.105 &
  0.192 &
  0.277 &
  0.356 &
  0.433 &
  0.111 &
  0.135 &
  0.228 &
  0.324 &
  0.408 &
  0.485 &
  0.174 &
  0.151 &
  0.309 &
  0.450 &
  0.565 &
  0.649 &
  0.125 
  \\
\oldvariant{}-sub$_{\textit{ngram}}$ &
  0.095 &
  0.166 &
  0.237 &
  0.311 &
  0.392 &
  0.081 &
  0.118 &
  0.193 &
  0.270 &
  0.347 &
  0.433 &
  0.118 &
  0.097 &
  0.181 &
  0.274 &
  0.381 &
  0.489 &
  0.075 \\
   \midrule
\tech{}$_{\textit{codebert}}$  &
  \textbf{0.066} &
  \textbf{0.126} &
  \textbf{0.185} &
  \textbf{0.240} &
  \textbf{0.309} &
  \textbf{0.099} &
  \textbf{0.064} &
  \textbf{0.131} &
  \textbf{0.190} &
  \textbf{0.255} &
  \textbf{0.331} &
  \textbf{0.130} &
  \textbf{0.115} &
  \textbf{0.206} &
  \textbf{0.290} &
  \textbf{0.362} &
  \textbf{0.440} &
  \textbf{0.128} 
  \\
\tech{}$_{\textit{codebert}}$ (n=2) &
  0.065 &
  0.121 &
  0.180 &
  0.239 &
  0.306 &
  0.096 &
  0.059 &
  0.117 &
  0.181 &
  0.251 &
  0.329 &
  0.126 &
  0.098 &
  0.182 &
  0.269 &
  0.347 &
  0.429 &
  0.112 \\
  \codebert{} &
  0.050 &
  0.089 &
  0.136 &
  0.188 &
  0.263 &
  0.081 &
  0.056 &
  0.102 &
  0.156 &
  0.215 &
  0.294 &
  0.114 &
  0.082 &
  0.147 &
  0.208 &
  0.285 &
  0.373 &
  0.107 \\
\oldvariant{}-sub$_{\textit{codebert}}$ &
  0.061 &
  0.116 &
  0.177 &
  0.236 &
  0.307 &
  0.086 &
  0.058 &
  0.112 &
  0.178 &
  0.245 &
  0.323 &
  0.089 &
  0.098 &
  0.184 &
  0.269 &
  0.347 &
  0.429 &
  0.105 \\
  \bottomrule
\end{tabular}
\end{threeparttable}
}
\end{table*}

Table~\ref{tab:rq3-prioritization} shows the comparison results in terms of average RAUC-k and MRR across all bugs in each dataset.
Due to the space limit, we put the detailed results on each bug by each method of measuring code naturalness at our project homepage.
From Table~\ref{tab:rq3-prioritization}, \tech{}$_{\textit{ngram}}$ outperforms \old{} and \tech{}$_{\textit{codebert}}$ outperforms \codebert{} on all three datasets in terms of all the metrics.
For example,
on average across all three datasets, \tech{}$_{\textit{ngram}}$/\tech{}$_{\textit{codebert}}$ achieves an improvement of 14.81\%$\sim$28.48\%/14.29\%$\sim$40.24\% over \old{}/\codebert{} in terms of RAUC-20\% 
and 2.30\%$\sim$6.31\%/14.04\%$\sim$22.22\% in terms of MRR.
In particular, the improvements of \tech{}$_{\textit{ngram}}$ over \old{} and \tech{}$_{\textit{codebert}}$ over \codebert{} are larger when $k$ is smaller, indicating that we can identify buggy lines more efficiently especially when the given inspection time is limited.
\revise{
The results also show that \tech{}$_{\textit{ngram}}$/\tech{}$_{\textit{codebert}}$ outperforms \oldvariant{}-sub$_{\textit{ngram}}$/\oldvariant{}-sub$_{\textit{codebert}}$ respectively in buggy line prioritization across all three datasets. 
For example, \tech{}$_{\textit{ngram}}$ and \tech{}$_{\textit{codebert}}$ improve upon their \oldvariant{}-sub counterparts by 31.36\%$\sim$100.00\% and 8.20\%$\sim$17.35\% respectively, across all three datasets in terms of RAUC-20\%. 
For different n values, the results demonstrate that DAN with n=3 consistently outperforms that with n=2, but the latter still significantly outperforms the baselines.
}
Overall, the results show the stable and obvious superiority of \tech{} over the baselines in identifying buggy lines among all code lines.

Furthermore, we performed a paired sample Wilcoxon signed-rank test~\cite{woolson2007wilcoxon} at the significance level of 0.05 to check whether each metric value achieved by \tech{}$_{\textit{ngram}}$/\tech{}$_{\textit{codebert}}$ is significantly better than that of \old{}/\codebert{} in statistics.
We found that all the p-values are smaller than 0.05 \revise{and the effect size between \tech{}$_{\textit{ngram}}$/\tech{}$_{\textit{codebert}}$ and Ngram-NT/CodeBERT-NT is 0.33/0.55 in terms of MRR} across the three datasets, further confirming the superiority of \tech{}. 

\begin{tcolorbox}
\textbf{Finding 2:} 
Incorporating code dependency for measuring code naturalness with \tech{} can more effectively identify buggy lines from all the code lines than the state-of-the-art \old{} and \codebert{}, demonstrating that \tech{} provides more accurate measure for code naturalness in the application of buggy line prioritization.

\end{tcolorbox}

\subsection{RQ3: Effectiveness Comparison on Training Data Cleansing}

Code generation holds great promise in automating various aspects of software development and the naturalness (such as readability and maintainability) of generated code is an important property (besides the widely-watched performance metric -- functionality correctness of generated code)~\cite{austin2021program}.
Cleansing training data with code naturalness measure (such as \tech{}) is a potential way of building the code generation models ensuring this property.
Moreover, the existing work has indicated that small-scale high-quality training data can achieve performance levels in code generation that are comparable to models trained on larger datasets~\cite{gunasekar2023textbooks}. 
Indeed, sampling training data with high code naturalness can also ensure the high quality of data to a large extent due to guaranteeing syntactic correctness as well, and thus such a data cleansing strategy will not damage code generation performance too much.
In this RQ, we investigated whether cleansing training data by emphasizing code naturalness with \tech{} (compared to the state-of-the-art \old{} and \codebert{}) can lead to better models that generate code with a more natural and readable structure without sacrificing overall code generation performance.

\subsubsection{Datasets}
To answer RQ3, our subjects include the models and datasets for code generation.
Here, we cleansed training data in each dataset, and then trained or fined-tuned each model based on the cleansed data, in order to build a better model.
In the study, we used two pre-trained code generation models, i.e., CodeGen-Multi~\cite{nijkamp2022codegen} and GPT-2~\cite{radford2019language}, as the target models due to the availability, popularity, and reasonable training cost. 
Moreover, we used APPS~\cite{hendrycks2021apps} and HumanEval-X
~\cite{zheng2023codegeex} as the datasets, which has been widely used in the task of code generation~\cite{le2022coderl,CERT,zhang2023planning,nijkamp2022codegen,zheng2023codegeex}.
HumanEval-X has been introduced in Section~\ref{sec:rq1}, for which we randomly sampled 90\% of data as training data and regarded the remaining data as test data.
APPS consists of 10,000 programming problems gathered from various open-access programming websites.
It has been officially divided into training and test data, each of which contains 5,000 programming problems.
These problems are categorized into three levels according to their degrees of difficulty. 
Following the existing work~\cite{olausson2023demystifying}, we randomly sampled 300 problems from the APPS test set according to the original distribution of the three levels, in order to balance evaluation cost and conclusion generality.
As the current implementation of \tech{} is for Java code, we used the Java versions of HumanEval-X and APPS.
The former is provided by the existing work~\cite{zheng2023codegeex}, while the latter is constructed by us with the aid of ChatGPT~\cite{openai_chatgpt}.
\revise{In the Java version of APPS, 57.95\% of statements are dependent on other statements and each dependency sequence involves 3 statements on average.}
We also released the Java version of APPS at our project homepage for future research.

\subsubsection{Process}
For each dataset, we first measured the naturalness of each training code snippet by each method, and ranked all the training code snippets in the ascending order of the measured naturalness scores.
Then, we selected top-50\% training data as the cleansed data, which are used to fine-tune each pre-trained model.
In this way, a fine-tuned model is built based on cleansed training data by each method of measuring code naturalness.

For sufficient comparison, we also used all training data in each dataset without cleansing to fine-tune each pre-trained model.
Besides, we included a baseline, which randomly selected 50\% training data from each dataset to fine-tune each pre-trained model.
In total, for a pair of dataset and pre-trained model, we obtained six fine-tuned models, among which two models are based on the cleansed training data by \old{} and \tech{}$_{\textit{ngram}}$, two models are based on the cleansed training data by \codebert{} and \tech{}$_{\textit{codebert}}$, one model is based on randomly sampled training data, and one model is based on all training data.
For the former five, the size of cleansed training data for model fine-tuning is the same (i.e., 50\% of the whole training data) for fair comparison.
The fine-tuning process follows the common practice in the existing work~\cite{yan2023coco,chen2021testing}. 
Finally, we compared the performance of these models on the test data in the corresponding dataset.
Note that we did not cleanse test data to avoid the overfitting issue.

\subsubsection{Metrics}

We measured the performance of fine-tuned models using two metrics: CodeBLEU~\cite{Ren2020CodeBLEU} and AvgPassRatio (Average Pass Ratio)~\cite{tian2023testcasedriven}. 
CodeBLEU measures the textual similarity between generated code and ground-truth code by considering both syntax match (the proportion of matched subtrees between their ASTs) and dataflow match (the proportion of matched def-use edges in their dataflow graphs).
Since the ground-truth code is meticulously crafted by developers, CodeBLEU can help measure the readability of generated code. 
That is, higher CodeBLEU scores mean that the generated code looks more similar to the developer-written ground truth, indicating higher readability of generated code to a large extent.
It is also possible that the generated code is significantly different from the ground-truth code but is also natural and readable.
For these cases, CodeBLEU could produce false negatives.
However, we did not find such cases by manually analyzing a small set of data via random sampling.
Indeed, CodeBLEU's effectiveness in measuring readability has been demonstrated by numerous studies~\cite{Ren2020CodeBLEU, yan2023coco, yang2024important}.
Therefore, the threat from CodeBLEU may be not serious.
Currently, it is still an open challenge to design an automated metric that can entirely eliminate false negatives in terms of readability~\cite{msra_survey}, which can be regarded as our future work.
Even though CodeBLEU has been also used to measure code generation performance, some recent work has pointed out its limitation in this aspect~\cite{roy2021reassessing}, and thus we just used it to measure code readability.

Besides, we used AvgPassRatio to measure code generation performance (the functionality correctness of generated code).
AvgPassRatio first calculates the ratio of passing test cases to all executed test cases on each piece of generated code, and then calculates the average ratio across the entire test set.
Larger AvgPassRatio values mean better code generation performance.

\subsubsection{Results and Analysis}
Table~\ref{tab:data-cleansing} 
shows the CodeBLEU score 
of each fine-tuned model based on different sets of (cleansed) training data for each pair of pre-trained model and dataset, respectively.
Row ``All'' presents the result of each fine-tuned model based on all training data in each dataset.
The last five rows present the results of each fine-tuned model based on the cleansed training data in each dataset by five different methods, respectively.
\begin{table}[t]
\caption{CodeBLEU effectiveness on training data cleansing}
\label{tab:data-cleansing}
\resizebox{0.5\linewidth}{!}{
\begin{tabular}{l|rr|rr}
\toprule
\multicolumn{1}{l|}{\multirow{2}{*}{\textbf{Method}}} &
  \multicolumn{2}{c|}{\textbf{HumanEval-X}} &
  \multicolumn{2}{c}{\textbf{APPS}} \\
\multicolumn{1}{l|}{} &
  \multicolumn{1}{c}{\textbf{CodeGen}} &
  \multicolumn{1}{c|}{\textbf{GPT-2}} &
  \multicolumn{1}{c}{\textbf{CodeGen}} &
  \multicolumn{1}{c}{\textbf{GPT-2}} \\
  \midrule
\multicolumn{1}{l|}{All}                                   & 0.43 & 0.14 & 0.34 & 0.33 \\
\multicolumn{1}{l|}{Random}                                & 0.40 & 0.07 & 0.33 & 0.32 \\
\midrule
\multicolumn{1}{l|}{\old{}}               & 0.37 & 0.17 & 0.34 & 0.32 \\
\multicolumn{1}{l|}{\tech{}$_{\textit{ngram}}$}    & 0.46 & 0.21 & 0.37 & 0.37 \\
\midrule
\multicolumn{1}{l|}{\codebert{}}          & 0.47 & 0.14 & 0.40   & 0.31 \\
\multicolumn{1}{l|}{\tech{}$_{\textit{codebert}}$} & 0.49 & 0.20 & 0.42   & 0.35 \\
\bottomrule
\end{tabular}
}
\end{table}

From Table~\ref{tab:data-cleansing}, the fine-tuned models with the aid of \tech{} can generate code with higher readability (measured by CodeBLEU) than those based on all training data, demonstrating the importance of cleansing training data for model building.
This aligns with the existing work~\cite{natgen, gunasekar2023textbooks}, reinforcing the conclusion that training with natural code can enhance the readability of generated code. 
However, random sampling cannot achieve better model performance than using all training data on all the subjects, highlighting the necessity of designing effective data cleansing methods.
Notably, \tech{}$_{\textit{ngram}}$ achieves an average improvement of 18.08\% over \old{} in facilitating generating natural code, and \tech{}$_{\textit{codebert}}$ achieves an average improvement of 16.25\% over \codebert{}, across all the subjects.

Furthermore, both \tech{}$_{\textit{ngram}}$ and \tech{}$_{\textit{codebert}}$ do not make their fine-tuned models sacrifice overall code generation performance.
In terms of AvgPassRatio, their fine-tuned models even slightly improve the code generation performance by 0.90\%$\sim$8.66\% over the fine-tuned models with all training data and 0.12\%$\sim$6.85\% over the fine-tuned models with randomly sampled training data.
However, both \old{} and \codebert{} make at least one fine-tuned models perform worse than fine-tuning with all training data or randomly sampled training data in terms of AvgPassRatio.
The results demonstrate the superiority of \tech{} over the state-of-the-are methods of measuring code naturalness in cleansing training data for building better models, which can generate more natural code without damaging the overall functionality correctness of the generated code.

\begin{tcolorbox}
\textbf{Finding 3:} 
Incorporating code dependency for measuring code naturalness with \tech{} can more effectively cleanse training data for building better models (that generate more natural code without sacrificing overall code generation performance), which outperforms the data cleansing methods with the state-of-the-art \old{} and \codebert{}, random sampling, and the way of using all training data.

\end{tcolorbox}

\del{
\subsection{RQ4: Contribution of Our Code Dependency Extraction}
\subsubsection{Process}
As presented before, 
we conducted comparisons among \tech{}$_{\textit{ngram}}$, \oldvariant{}$_{\textit{ngram}}$, and \oldvariant{}-sub$_{\textit{ngram}}$, as well as between \tech{}$_{\textit{codebert}}$ and \oldvariant{}-sub$_{\textit{codebert}}$, 
to investigate the contribution of the code dependency extraction strategy in \tech{}.
As fine-tuning code models is resource-intensive and time-consuming, we answered this RQ in the first two scenarios (i.e., distinguishing natural and unnatural code, and buggy line prioritization).
}

\del{
\subsubsection{Results and Analysis}
Table~\ref{tab:rq2-mutation-1} shows the comparison results on distinguishing natural and unnatural code.
The results show that \tech{}$_{\textit{ngram}}$ outperforms \oldvariant{}-sub$_{\textit{ngram}}$ and \tech{}$_{\textit{codebert}}$ outperforms \oldvariant{}-sub$_{\textit{codebert}}$ in the vast majority of cases.
On average, \tech{}$_{\textit{ngram}}$ shows an improvement of 58.65\% over \oldvariant{}-sub$_{\textit{ngram}}$, and \tech{}$_{\textit{codebert}}$ shows an average improvement of 136.75\% over \oldvariant{}-sub$_{\textit{codebert}}$,  in terms of average normalized difference values.
In additional, \tech{}$_{\textit{ngram}}$ always performs better than \oldvariant{}$_{\textit{ngram}}$.
}

\del{
Table~\ref{tab:rq3-prioritization} shows the results on buggy line prioritization. 
The results show that \tech{}$_{\textit{ngram}}$/\tech{}$_{\textit{codebert}}$ outperforms \oldvariant{}-sub$_{\textit{ngram}}$/\oldvariant{}-sub$_{\textit{codebert}}$ respectively in buggy line prioritization across all three datasets. 
For example, \tech{}$_{\textit{ngram}}$ and \tech{}$_{\textit{codebert}}$ improve upon their \oldvariant{}-sub counterparts by 31.36\%$\sim$100.00\% and 8.20\%$\sim$17.35\% respectively, across all three datasets in terms of RAUC-20\%. 
}

\del{
The results not only confirm the contribution of relatively broad dependency information compared to the local context information incorporated by \oldvariant{}-sub, but also confirm the poor performance of directly treating the code as a whole sequence due to incurring too much noise (even though including all dependency information).
It demonstrates the importance of incorporating code dependency in a reasonable manner for measuring code naturalness (like \tech{}).
}

\del{
\begin{tcolorbox}
\textbf{Finding 4:} 
Our code dependency extraction strategy contribute to the overall effectiveness of \tech{} compared to the intuitive methods of treating code as a whole sequence to include dependency information (regardless of using subsequence extraction or not).
\end{tcolorbox}
}

\section{Discussion}
\label{sec:discussion}
\subsection{Contribution of Sub-path Extraction}
\begin{table}[t]
\renewcommand\arraystretch{1.2} 
\caption{Effectiveness comparison between \tech{} and its variant w/o sub path (SP) in distinguishing unnatural code}
\label{tab:discussion-1}
\resizebox{0.5\linewidth}{!}{
\begin{tabular}{l|c|c}
\toprule
\multicolumn{1}{c|}{\textbf{Transformation Rules}} &
  \multicolumn{1}{c|}{\textbf{\tech{}$_{ngram}$ w/o SP}} &
  \multicolumn{1}{c}{\textbf{\tech{}$_{ngram}$}} 
  \\
  \midrule
Dead Code Insertion &
  12.50\% &
  \textbf{20.26\%} 
  \\
Confusing Code Insertion &
  24.99\% &
  \textbf{40.94\%} 
  \\
Variable Renaming &
  33.54\% &
  \textbf{34.63\%} 
  \\
  \bottomrule
\end{tabular}
}
\end{table}

As presented in Section~\ref{sec:approach}, one important component in \tech{} is to extract n-node sub-paths from each complete path for facilitating the measure of code naturalness.
Here, we conducted an experiment to empirically evaluate the contribution of this component by comparing \tech{}$_{\textit{ngram}}$ with its variant without this component.
Specifically, this variant transforms a complete path (instead of a sub-path) into a sequence of code lines for naturalness measure.
We repeated the experiment of RQ1 based on \tech{}$_{\textit{ngram}}$ and this variant.
Table~\ref{tab:discussion-1} presents the comparison results between them in terms of the average normalized difference values.
The results demonstrate that \tech{}$_{\textit{ngram}}$ significantly outperforms the variant of \tech{}$_{\textit{ngram}}$ without the sub-path extraction component in distinguishing natural code and unnatural code.
The average improvement of \tech{}$_{\textit{ngram}}$ over this variant is 56.59\% in terms of average normalized difference values across all the 12 cases.
The results confirm the contribution of the sub-path extraction component in \tech{}$_{\textit{ngram}}$.

\subsection{Effectiveness of \tech{} with More Advanced Statistical Models}
\label{sec:llms}

\begin{table}[t]
\renewcommand\arraystretch{1.2} 
\caption{Effectiveness of \tech{} with more advanced statistical models}
\label{tab:discussion-2}
\resizebox{0.5\linewidth}{!}{
\begin{tabular}{l|c|c}
\toprule
\multicolumn{1}{c|}{\textbf{Transformation Rules}} &
  \textbf{CodeLlama} &
  \textbf{\tech{}$_{CodeLlama}$} 
  \\
  \midrule
Dead Code Insertion &
  2.64\% &
  \textbf{5.19\%}
  \\
Confusing Code Insertion &
    5.32\% &
  \textbf{7.09\%}
  \\
Variable Renaming &
    2.85\% &
  \textbf{10.83\%}
  \\
  \bottomrule
\end{tabular}
}
\end{table}

As discussed before, our \tech{} idea can be generally applied independently of the used statistical model. 
In our study, we chose the cached n-gram model and the pre-trained CodeBERT model as the underlying statistical model in \tech{} for evaluation, respectively.
This is because there are existing studies that proposed the corresponding methods of measuring code naturalness using the two models.
Here, we further investigated whether integrating a more advanced large language model in \tech{} can also outperform the corresponding method of directly using the large language model to measure code naturalness of each line, which can be helpful to further confirm the generalizability of \tech{} and the orthogonality between our \tech{} idea and the underlying statistical models.
Specifically, we used the CodeLlama-13b-hf model~\cite{codellama} as the representative large language model for investigation.
Accordingly, we constructed the instantiation \tech{}$_{\textit{CodeLlama}}$ and the baseline CodeLlama-NT (which employs CodeLlama-13b-hf to measure code naturalness of each line individually).
We repeated the experiment of RQ1 based on \tech{}$_{\textit{CodeLlama}}$ and CodeLlama-NT.
Table~\ref{tab:discussion-2} presents the comparison results between them in terms of the average normalized difference values.
The results demonstrate that \tech{}$_{\textit{CodeLlama}}$ outperforms CodeLlama-NT in distinguishing natural and unnatural code, achieving an average improvement of 1.24 times across all 12 cases in terms of average normalized difference values.
These results further confirm the generalizability of \tech{}, demonstrating its consistent enhancement on code naturalness measure regardless of the underlying statistical models.

\subsection{Future Work}
\revise{
Although \tech{} does not necessarily achieve state-of-the-art performance on the two downstream applications as our primary goal is to demonstrate the superiority of \tech{}$_{\textit{ngram}}$ over \old{} and \tech{}$_{\textit{codebert}}$ over \codebert{}, future work could explore how \tech{} can be integrated with state-of-the-art methods to enhance their effectiveness.
For example, in the scenario of buggy line prioritization, spectrum-based fault localization (SBFL) is a state-of-the-art method for buggy line prioritization based on test coverage. However, SBFL suffers from the tie problem~\cite{li2019deepfl}, where lines within the same block are given the same priority. \tech{} could address this issue by distinguishing between lines that belong to different sub-paths, thereby breaking ties in SBFL and potentially improving its accuracy.
In the scenario of training data cleansing, code evaluation involves multiple dimensions, such as syntactic validity, semantic correctness, and naturalness. While state-of-the-art cleansing methods focus primarily on validity and diversity~\cite{gunasekar2023textbooks, guo2024deepseek}, \tech{} offers a complementary approach by enhancing code naturalness, which could lead to more comprehensive improvements in code-generation performance.

Moreover, the model-agnostic nature of \tech{} presents opportunities for further exploration across a broader range of models and tasks, such as code recommendation or code retrieval. \tech{} is also suitable to distinguish LLM-generated and human-written code as a more precise measure in code naturalness, since the recent work~\cite{xu2024detecting} indicates that slight alterations to LLM-generated code typically cause higher perplexity to the model, making the code appear more unnatural. This further exploration could reveal whether \tech{} can serve as a static analysis way to address nonlinearities in tasks involving code, in contrast to approaches that rely on learning nonlinear context.

Currently, \tech{} depends on PDGs, which have scalability limitations. Future research could focus on alleviating scalability challenges. Additionally, \tech{} has been built on the assumption that naturalness has to be line-level, and implemented on Java, with some experiments conducted on the relatively small HumanEval dataset. In the future, we plan to evaluate \tech{} on a more diverse set of subjects, including different granularities, programming languages, and datasets.
}

\subsection{Threats to Validity}
\label{sec:threats}
The \textit{internal} threat mainly arises from the implementation of \tech{} and the compared techniques. 
To reduce this threat, we conducted thorough code review and replicated the original results of \old{} in its original study based on our re-implementation~\cite{buggy_code_naturalness}.

The threat to \textit{external} validity mainly lies in the subjects used. 
As presented in Section~\ref{sec:evaluation}, we implemented and evaluated \tech{} on Java code, which may not represent the subjects under other programming languages. 
In fact, the idea of \tech{} is general and can be implemented to the code written in other programming languages, as long as there is a program analysis tool supporting the extraction of code dependency as \tech{}.
In the future, we will extend and evaluate \tech{} on a more diverse set of subjects, encompassing different programming languages, to reduce this threat.

The threats to \textit{construct} validity mainly lie in the randomness involved in our study, i.e., the construction of unnatural code in RQ1 and the baseline of randomly sampling training data in RQ3, the setting of $n$ in the sub-path extraction component, \revise{and the metrics.}
For the first kind of randomness, we used each transformation rule to construct four pieces of unnatural code with different degrees of de-naturalization from each piece of natural code. 
Such an extensive dataset can help reduce this threat.
For the second kind of randomness, we repeated the experiment five times and calculated the average results.
For the second threat, we set $n$ to 3 following the existing study~\cite{rahman2019natural_revisited}.
In the future, we will investigate the influence of this parameter to further reduce this threat.
\revise{
For the last threat, we have employed the widely-used metrics by following previous studies~\cite{chen2023exploring, wang2021prioritizing, yang2023silent, Ren2020CodeBLEU, tian2023testcasedriven}. 
In fact, we also analyzed the overhead of our technique by taking Defects4j (real-world Java projects) as the representative benchmark. 
On average, the time spent on extracting dependencies and sub-paths by \tech{} is just 3s per file. 
The time spent on measuring naturalness per file by DAN$_{ngram}$/DAN$_{codebert}$/Ngram-NT/CodeBERT-NT is 0.0099s/0.0283s/0.0049s/0.0209s. 
Overall, the overhead of \tech{} is acceptable especially when considering its significant effectiveness. 

}

\section{Related Work}

In the literature, there are some existing methods of measuring code naturalness~\cite{buggy_code_naturalness,software_naturalness,tu2014localness}, which are very relevant to our work.
Besides the state-of-the-art \old{}~\cite{buggy_code_naturalness} and \codebert{}~\cite{khanfir2022codebert-nt} that have been presented in Section~\ref{sec:background} and compared in our study,
there are some other methods, upon which \old{} are designed. 
For example, Hindle et al.~\cite{software_naturalness} applied the n-gram model and cross entropy to measure code naturalness. Tu et al. ~\cite{tu2014localness} incorporated a cache model to exploit localness of source code for measuring code naturalness more precisely.
Our work goes beyond the per-line measure for code naturalness and incorporates code dependency to improve the precision of measuring code naturalness.

Code naturalness has been utilized for several software engineering applications~\cite{buggy_code_naturalness, jit, tu2014localness}, such as buggy line prioritization and training data cleansing as evaluated in Section~\ref{sec:evaluation}.
Besides, Raychev et al.~\cite{2014Completion} used the naturalness of code APIs to build the language model for code completion.
Allamanis et al.~\cite{2014convention} learned the style of a codebase and utilized software naturalness to suggest revisions for improving stylistic consistency.
Dantas et al.~\cite{dantas2019code} proposed to assist search space exploration in program repair using code naturalness.
YAN et al.~\cite{jit} proposed a just-in-time defect prediction framework based on code naturalness.
Different from them, our work investigated the contribution of incorporating code dependency for code naturalness measure by designing \tech{} and improved the effectiveness of several naturalness applications.
In the future, we will improve the effectiveness of more naturalness applications based on \tech{}.

There are also some empirical studies on code naturalness~\cite{rahman2019natural_revisited,buggy_revisited}.
Rahman et al.~\cite{rahman2019natural_revisited} conducted an investigation into the impact of syntax tokens on naturalness measure and explored how different code representations exhibit different levels of repetition, which prompts a suggestion for future research to concentrate on new code representations for code naturalness.
However, they did not involve developing new methods to improve the measure of code naturalness, which contrasts with our method of utilizing code dependency to achieve it. 
Jiang et al.~\cite{buggy_revisited} conducted an extensive study to investigate the naturalness of buggy, non-buggy, and bug-fixing code, and confirmed that buggy code are less natural.
Different from this work, we investigated whether incorporating code dependency can improve the measure of code naturalness and thus facilitate the downstream applications.

Some studies have looked at leveraging code dependencies for the tasks related to code embedding, such as the pre-trained GraphCodeBERT~\cite{guo2020graphcodebert} and GraphCode2Vec models~\cite{10.1145/3524842.3528456}.
Different from them, our work aims to investigate the influence of code dependency on code naturalness measure, which can be regarded as a basis for these tasks.

Code naturalness is also related to code smell, since there is often a correlation between lower code naturalness and the presence of code smell. 
Code that is less natural tends to be harder to understand and maintain, which can lead to the introduction of code smells. 
For example, complex or convoluted code structures, poor naming conventions, and inconsistent formatting can all contribute to decreased naturalness and increase the likelihood of code smells. 
Conversely, code that exhibits higher naturalness, with clear and concise structure, meaningful variable names, and consistent formatting, is generally less prone to containing code smells. 
However, it is important to note that while there is a correlation, it is not a strict rule, code may lack naturalness due to factors such as complex logic, tight constraints, or specialized requirements, rather than due to inherent issues like code smells.
Conversely, code that appears natural and readable may still contain code smells if certain design or implementation choices introduce inefficiencies or make maintenance difficult.
In essence, while naturalness can be an indicator of code quality, it does not guarantee the absence or presence of code smell, and vice versa. Both aspects should be considered independently.

\section{Conclusion}
In this work, we have conducted the first empirical study to investigate whether incorporating code dependency, instead of isolating each code line like existing methods, can improve the precision of measuring code naturalness.
To achieve that, we first proposed a reasonable yet general method named \tech{} for incorporating the rich code dependency information in measuring code naturalness. 
The evaluation results of \tech{} in three emerging applications of code naturalness demonstrate that the code dependency information is significant in measuring code naturalness, encouraging more future research on it.

\section*{Data-Availability Statement}
We released our artifact and experimental data for replication and practical use~\cite{chenyang_2024_12783666}.

\begin{acks}
We thank the anonymous reviewers for their valuable suggestions. 
The work has been supported by the National Natural Science Foundation of China Grant Nos. 62322208, 12411530122, 62232001, 62202324, CCF Young Elite Scientists Sponsorship Program (by CAST).
\end{acks}


\bibliographystyle{ACM-Reference-Format}
\bibliography{reference}


\begin{thebibliography}{63}


\ifx \showCODEN    \undefined \def \showCODEN     #1{\unskip}     \fi
\ifx \showDOI      \undefined \def \showDOI       #1{#1}\fi
\ifx \showISBNx    \undefined \def \showISBNx     #1{\unskip}     \fi
\ifx \showISBNxiii \undefined \def \showISBNxiii  #1{\unskip}     \fi
\ifx \showISSN     \undefined \def \showISSN      #1{\unskip}     \fi
\ifx \showLCCN     \undefined \def \showLCCN      #1{\unskip}     \fi
\ifx \shownote     \undefined \def \shownote      #1{#1}          \fi
\ifx \showarticletitle \undefined \def \showarticletitle #1{#1}   \fi
\ifx \showURL      \undefined \def \showURL       {\relax}        \fi
\providecommand\bibfield[2]{#2}
\providecommand\bibinfo[2]{#2}
\providecommand\natexlab[1]{#1}
\providecommand\showeprint[2][]{arXiv:#2}

\bibitem[gro(2023)]%
        {growing}
 \bibinfo{year}{Accessed: 2023}\natexlab{}.
\newblock \bibinfo{title}{GrowingBugs}.
\newblock
\newblock
\newblock
\shownote{\url{https://github.com/jiangyanjie/GrowingBugs/}}.


\bibitem[jav(2023)]%
        {javalang}
 \bibinfo{year}{Accessed: 2023}\natexlab{}.
\newblock \bibinfo{title}{javalang}.
\newblock
\newblock
\newblock
\shownote{\url{https://github.com/c2nes/javalang/}}.


\bibitem[ope(2023)]%
        {openai_chatgpt}
 \bibinfo{year}{Accessed: 2023}\natexlab{}.
\newblock \bibinfo{title}{OpenAI ChatGPT}.
\newblock
\newblock
\newblock
\shownote{\url{https://chat.openai.com/}}.


\bibitem[skl(2023)]%
        {sklearn}
 \bibinfo{year}{Accessed: 2023}\natexlab{}.
\newblock \bibinfo{title}{scikit-learn}.
\newblock
\newblock
\newblock
\shownote{\url{https://scikit-learn.org/stable/}}.


\bibitem[Allamanis et~al\mbox{.}(2014)]%
        {2014convention}
\bibfield{author}{\bibinfo{person}{Miltiadis Allamanis}, \bibinfo{person}{Earl~T Barr}, \bibinfo{person}{Christian Bird}, {and} \bibinfo{person}{Charles Sutton}.} \bibinfo{year}{2014}\natexlab{}.
\newblock \showarticletitle{Learning Natural Coding Conventions}.
\newblock \bibinfo{journal}{\emph{ACM}} (\bibinfo{year}{2014}).
\newblock


\bibitem[Athiwaratkun et~al\mbox{.}(2022)]%
        {athiwaratkun2022multi}
\bibfield{author}{\bibinfo{person}{Ben Athiwaratkun}, \bibinfo{person}{Sanjay~Krishna Gouda}, \bibinfo{person}{Zijian Wang}, \bibinfo{person}{Xiaopeng Li}, \bibinfo{person}{Yuchen Tian}, \bibinfo{person}{Ming Tan}, \bibinfo{person}{Wasi~Uddin Ahmad}, \bibinfo{person}{Shiqi Wang}, \bibinfo{person}{Qing Sun}, \bibinfo{person}{Mingyue Shang}, {et~al\mbox{.}}} \bibinfo{year}{2022}\natexlab{}.
\newblock \showarticletitle{Multi-lingual Evaluation of Code Generation Models}. In \bibinfo{booktitle}{\emph{The Eleventh International Conference on Learning Representations}}.
\newblock


\bibitem[Austin et~al\mbox{.}(2021)]%
        {austin2021program}
\bibfield{author}{\bibinfo{person}{Jacob Austin}, \bibinfo{person}{Augustus Odena}, \bibinfo{person}{Maxwell Nye}, \bibinfo{person}{Maarten Bosma}, \bibinfo{person}{Henryk Michalewski}, \bibinfo{person}{David Dohan}, \bibinfo{person}{Ellen Jiang}, \bibinfo{person}{Carrie Cai}, \bibinfo{person}{Michael Terry}, \bibinfo{person}{Quoc Le}, {et~al\mbox{.}}} \bibinfo{year}{2021}\natexlab{}.
\newblock \showarticletitle{Program synthesis with large language models}.
\newblock \bibinfo{journal}{\emph{arXiv preprint arXiv:2108.07732}} (\bibinfo{year}{2021}).
\newblock


\bibitem[Busjahn et~al\mbox{.}(2015)]%
        {busjahn2015eye}
\bibfield{author}{\bibinfo{person}{Teresa Busjahn}, \bibinfo{person}{Roman Bednarik}, \bibinfo{person}{Andrew Begel}, \bibinfo{person}{Martha Crosby}, \bibinfo{person}{James~H Paterson}, \bibinfo{person}{Carsten Schulte}, \bibinfo{person}{Bonita Sharif}, {and} \bibinfo{person}{Sascha Tamm}.} \bibinfo{year}{2015}\natexlab{}.
\newblock \showarticletitle{Eye movements in code reading: Relaxing the linear order}. In \bibinfo{booktitle}{\emph{2015 IEEE 23rd International Conference on Program Comprehension}}. IEEE, \bibinfo{pages}{255--265}.
\newblock


\bibitem[Casalnuovo et~al\mbox{.}(2020a)]%
        {casalnuovo2020programmers}
\bibfield{author}{\bibinfo{person}{Casey Casalnuovo}, \bibinfo{person}{Kevin Lee}, \bibinfo{person}{Hulin Wang}, \bibinfo{person}{Prem Devanbu}, {and} \bibinfo{person}{Emily Morgan}.} \bibinfo{year}{2020}\natexlab{a}.
\newblock \showarticletitle{Do programmers prefer predictable expressions in code?}
\newblock \bibinfo{journal}{\emph{Cognitive science}} \bibinfo{volume}{44}, \bibinfo{number}{12} (\bibinfo{year}{2020}), \bibinfo{pages}{e12921}.
\newblock


\bibitem[Casalnuovo et~al\mbox{.}(2020b)]%
        {casalnuovo2020does}
\bibfield{author}{\bibinfo{person}{Casey Casalnuovo}, \bibinfo{person}{E Morgan}, {and} \bibinfo{person}{P Devanbu}.} \bibinfo{year}{2020}\natexlab{b}.
\newblock \showarticletitle{Does surprisal predict code comprehension difficulty}. In \bibinfo{booktitle}{\emph{Proceedings of the 42nd Annual Meeting of the Cognitive Science Society}}. Cognitive Science Society Toronto, Canada.
\newblock


\bibitem[Cavnar et~al\mbox{.}(1994)]%
        {cavnar1994n}
\bibfield{author}{\bibinfo{person}{William~B Cavnar}, \bibinfo{person}{John~M Trenkle}, {et~al\mbox{.}}} \bibinfo{year}{1994}\natexlab{}.
\newblock \showarticletitle{N-gram-based text categorization}. In \bibinfo{booktitle}{\emph{Proceedings of SDAIR-94, 3rd annual symposium on document analysis and information retrieval}}, Vol.~\bibinfo{volume}{161175}. Las Vegas, NV, \bibinfo{pages}{14}.
\newblock


\bibitem[Chakraborty et~al\mbox{.}(2022)]%
        {natgen}
\bibfield{author}{\bibinfo{person}{Saikat Chakraborty}, \bibinfo{person}{Toufique Ahmed}, \bibinfo{person}{Yangruibo Ding}, \bibinfo{person}{Premkumar~T Devanbu}, {and} \bibinfo{person}{Baishakhi Ray}.} \bibinfo{year}{2022}\natexlab{}.
\newblock \showarticletitle{NatGen: generative pre-training by “naturalizing” source code}. In \bibinfo{booktitle}{\emph{Proceedings of the 30th ACM Joint European Software Engineering Conference and Symposium on the Foundations of Software Engineering}}. \bibinfo{pages}{18--30}.
\newblock


\bibitem[Chen et~al\mbox{.}(2021b)]%
        {chen2021evaluating}
\bibfield{author}{\bibinfo{person}{Mark Chen}, \bibinfo{person}{Jerry Tworek}, \bibinfo{person}{Heewoo Jun}, \bibinfo{person}{Qiming Yuan}, \bibinfo{person}{Henrique~Ponde de Oliveira~Pinto}, \bibinfo{person}{Jared Kaplan}, \bibinfo{person}{Harri Edwards}, \bibinfo{person}{Yuri Burda}, \bibinfo{person}{Nicholas Joseph}, \bibinfo{person}{Greg Brockman}, {et~al\mbox{.}}} \bibinfo{year}{2021}\natexlab{b}.
\newblock \showarticletitle{Evaluating large language models trained on code.(2021)}.
\newblock \bibinfo{journal}{\emph{arXiv preprint arXiv:2107.03374}} (\bibinfo{year}{2021}).
\newblock


\bibitem[Chen et~al\mbox{.}(2021a)]%
        {chen2021testing}
\bibfield{author}{\bibinfo{person}{Songqiang Chen}, \bibinfo{person}{Shuo Jin}, {and} \bibinfo{person}{Xiaoyuan Xie}.} \bibinfo{year}{2021}\natexlab{a}.
\newblock \showarticletitle{Testing your question answering software via asking recursively}. In \bibinfo{booktitle}{\emph{2021 36th IEEE/ACM International Conference on Automated Software Engineering (ASE)}}. IEEE, \bibinfo{pages}{104--116}.
\newblock


\bibitem[Chen et~al\mbox{.}(2023)]%
        {chen2023exploring}
\bibfield{author}{\bibinfo{person}{Zhichao Chen}, \bibinfo{person}{Junjie Chen}, \bibinfo{person}{Weijing Wang}, \bibinfo{person}{Jianyi Zhou}, \bibinfo{person}{Meng Wang}, \bibinfo{person}{Xiang Chen}, \bibinfo{person}{Shan Zhou}, {and} \bibinfo{person}{Jianmin Wang}.} \bibinfo{year}{2023}\natexlab{}.
\newblock \showarticletitle{Exploring better black-Box test case prioritization via log analysis}.
\newblock \bibinfo{journal}{\emph{ACM Transactions on Software Engineering and Methodology}} \bibinfo{volume}{32}, \bibinfo{number}{3} (\bibinfo{year}{2023}), \bibinfo{pages}{1--32}.
\newblock


\bibitem[Dantas et~al\mbox{.}(2019)]%
        {dantas2019code}
\bibfield{author}{\bibinfo{person}{Altino Dantas}, \bibinfo{person}{Eduardo~F de Souza}, \bibinfo{person}{Jerffeson Souza}, {and} \bibinfo{person}{Celso~G Camilo-Junior}.} \bibinfo{year}{2019}\natexlab{}.
\newblock \showarticletitle{Code naturalness to assist search space exploration in search-based program repair methods}. In \bibinfo{booktitle}{\emph{Search-Based Software Engineering: 11th International Symposium, SSBSE 2019, Tallinn, Estonia, August 31--September 1, 2019, Proceedings 11}}. Springer, \bibinfo{pages}{164--170}.
\newblock


\bibitem[Feng et~al\mbox{.}(2020)]%
        {feng2020codebert}
\bibfield{author}{\bibinfo{person}{Zhangyin Feng}, \bibinfo{person}{Daya Guo}, \bibinfo{person}{Duyu Tang}, \bibinfo{person}{Nan Duan}, \bibinfo{person}{Xiaocheng Feng}, \bibinfo{person}{Ming Gong}, \bibinfo{person}{Linjun Shou}, \bibinfo{person}{Bing Qin}, \bibinfo{person}{Ting Liu}, \bibinfo{person}{Daxin Jiang}, {et~al\mbox{.}}} \bibinfo{year}{2020}\natexlab{}.
\newblock \showarticletitle{CodeBERT: A Pre-Trained Model for Programming and Natural Languages}. In \bibinfo{booktitle}{\emph{Findings of the Association for Computational Linguistics: EMNLP 2020}}. \bibinfo{pages}{1536--1547}.
\newblock


\bibitem[Gunasekar et~al\mbox{.}(2023)]%
        {gunasekar2023textbooks}
\bibfield{author}{\bibinfo{person}{Suriya Gunasekar}, \bibinfo{person}{Yi Zhang}, \bibinfo{person}{Jyoti Aneja}, \bibinfo{person}{Caio C{\'e}sar~Teodoro Mendes}, \bibinfo{person}{Allie Del~Giorno}, \bibinfo{person}{Sivakanth Gopi}, \bibinfo{person}{Mojan Javaheripi}, \bibinfo{person}{Piero Kauffmann}, \bibinfo{person}{Gustavo de Rosa}, \bibinfo{person}{Olli Saarikivi}, {et~al\mbox{.}}} \bibinfo{year}{2023}\natexlab{}.
\newblock \showarticletitle{Textbooks Are All You Need}.
\newblock \bibinfo{journal}{\emph{arXiv preprint arXiv:2306.11644}} (\bibinfo{year}{2023}).
\newblock


\bibitem[Guo et~al\mbox{.}(2020)]%
        {guo2020graphcodebert}
\bibfield{author}{\bibinfo{person}{Daya Guo}, \bibinfo{person}{Shuo Ren}, \bibinfo{person}{Shuai Lu}, \bibinfo{person}{Zhangyin Feng}, \bibinfo{person}{Duyu Tang}, \bibinfo{person}{LIU Shujie}, \bibinfo{person}{Long Zhou}, \bibinfo{person}{Nan Duan}, \bibinfo{person}{Alexey Svyatkovskiy}, \bibinfo{person}{Shengyu Fu}, {et~al\mbox{.}}} \bibinfo{year}{2020}\natexlab{}.
\newblock \showarticletitle{GraphCodeBERT: Pre-training Code Representations with Data Flow}. In \bibinfo{booktitle}{\emph{International Conference on Learning Representations}}.
\newblock


\bibitem[Guo et~al\mbox{.}(2024)]%
        {guo2024deepseek}
\bibfield{author}{\bibinfo{person}{Daya Guo}, \bibinfo{person}{Qihao Zhu}, \bibinfo{person}{Dejian Yang}, \bibinfo{person}{Zhenda Xie}, \bibinfo{person}{Kai Dong}, \bibinfo{person}{Wentao Zhang}, \bibinfo{person}{Guanting Chen}, \bibinfo{person}{Xiao Bi}, \bibinfo{person}{Y Wu}, \bibinfo{person}{YK Li}, {et~al\mbox{.}}} \bibinfo{year}{2024}\natexlab{}.
\newblock \showarticletitle{DeepSeek-Coder: When the Large Language Model Meets Programming--The Rise of Code Intelligence}.
\newblock \bibinfo{journal}{\emph{arXiv preprint arXiv:2401.14196}} (\bibinfo{year}{2024}).
\newblock


\bibitem[Hendrycks et~al\mbox{.}(2021)]%
        {hendrycks2021apps}
\bibfield{author}{\bibinfo{person}{Dan Hendrycks}, \bibinfo{person}{Steven Basart}, \bibinfo{person}{Saurav Kadavath}, \bibinfo{person}{Mantas Mazeika}, \bibinfo{person}{Akul Arora}, \bibinfo{person}{Ethan Guo}, \bibinfo{person}{Collin Burns}, \bibinfo{person}{Samir Puranik}, \bibinfo{person}{Horace He}, \bibinfo{person}{Dawn Song}, {et~al\mbox{.}}} \bibinfo{year}{2021}\natexlab{}.
\newblock \showarticletitle{Measuring coding challenge competence with apps}.
\newblock \bibinfo{journal}{\emph{arXiv preprint arXiv:2105.09938}} (\bibinfo{year}{2021}).
\newblock


\bibitem[Herbold et~al\mbox{.}(2020)]%
        {Herbold2020smartshark}
\bibfield{author}{\bibinfo{person}{Steffen Herbold}, \bibinfo{person}{Alexander Trautsch}, \bibinfo{person}{Benjamin Ledel}, \bibinfo{person}{Alireza Aghamohammadi}, \bibinfo{person}{Taher~Ahmed Ghaleb}, \bibinfo{person}{Kuljit~Kaur Chahal}, \bibinfo{person}{Tim Bossenmaier}, \bibinfo{person}{Bhaveet Nagaria}, \bibinfo{person}{Philip Makedonski}, \bibinfo{person}{Matin~Nili Ahmadabadi}, \bibinfo{person}{Kristof Szabados}, \bibinfo{person}{Helge Spieker}, \bibinfo{person}{Matej Madeja}, \bibinfo{person}{Nathaniel Hoy}, \bibinfo{person}{Valentina Lenarduzzi}, \bibinfo{person}{Shangwen Wang}, \bibinfo{person}{Gema Rodríguez-Pérez}, \bibinfo{person}{Ricardo Colomo-Palacios}, \bibinfo{person}{Roberto Verdecchia}, \bibinfo{person}{Paramvir Singh}, \bibinfo{person}{Yihao Qin}, \bibinfo{person}{Debasish Chakroborti}, \bibinfo{person}{Willard Davis}, \bibinfo{person}{Vijay Walunj}, \bibinfo{person}{Hongjun Wu}, \bibinfo{person}{Diego Marcilio}, \bibinfo{person}{Omar Alam}, \bibinfo{person}{Abdullah Aldaeej},
  \bibinfo{person}{Idan Amit}, \bibinfo{person}{Burak Turhan}, \bibinfo{person}{Simon Eismann}, \bibinfo{person}{Anna-Katharina Wickert}, \bibinfo{person}{Ivano Malavolta}, \bibinfo{person}{Matus Sulir}, \bibinfo{person}{Fatemeh Fard}, \bibinfo{person}{Austin~Z. Henley}, \bibinfo{person}{Stratos Kourtzanidis}, \bibinfo{person}{Eray Tuzun}, \bibinfo{person}{Christoph Treude}, \bibinfo{person}{Simin~Maleki Shamasbi}, \bibinfo{person}{Ivan Pashchenko}, \bibinfo{person}{Marvin Wyrich}, \bibinfo{person}{James Davis}, \bibinfo{person}{Alexander Serebrenik}, \bibinfo{person}{Ella Albrecht}, \bibinfo{person}{Ethem~Utku Aktas}, \bibinfo{person}{Daniel Strüber}, {and} \bibinfo{person}{Johannes Erbel}.} \bibinfo{year}{2020}\natexlab{}.
\newblock \bibinfo{title}{Large-Scale Manual Validation of Bug Fixing Commits: A Fine-grained Analysis of Tangling}.
\newblock
\newblock
\showeprint[arxiv]{2011.06244}~[cs.SE]


\bibitem[Hindle et~al\mbox{.}(2012)]%
        {software_naturalness}
\bibfield{author}{\bibinfo{person}{Abram Hindle}, \bibinfo{person}{Earl~T Barr}, \bibinfo{person}{Zhendong Su}, \bibinfo{person}{Mark Gabel}, {and} \bibinfo{person}{Premkumar Devanbu}.} \bibinfo{year}{2012}\natexlab{}.
\newblock \showarticletitle{On the naturalness of software}. In \bibinfo{booktitle}{\emph{2012 34th International Conference on Software Engineering (ICSE)}}. IEEE.
\newblock


\bibitem[Jiang et~al\mbox{.}(2022)]%
        {buggy_revisited}
\bibfield{author}{\bibinfo{person}{Yanjie Jiang}, \bibinfo{person}{Hui Liu}, \bibinfo{person}{Yuxia Zhang}, \bibinfo{person}{Weixing Ji}, \bibinfo{person}{Hao Zhong}, {and} \bibinfo{person}{Lu Zhang}.} \bibinfo{year}{2022}\natexlab{}.
\newblock \showarticletitle{Do bugs lead to unnaturalness of source code?}. In \bibinfo{booktitle}{\emph{Proceedings of the 30th ACM Joint European Software Engineering Conference and Symposium on the Foundations of Software Engineering}}. \bibinfo{pages}{1085--1096}.
\newblock


\bibitem[Just et~al\mbox{.}(2014)]%
        {just2014defects4j}
\bibfield{author}{\bibinfo{person}{Ren{\'e} Just}, \bibinfo{person}{Darioush Jalali}, {and} \bibinfo{person}{Michael~D Ernst}.} \bibinfo{year}{2014}\natexlab{}.
\newblock \showarticletitle{Defects4J: A database of existing faults to enable controlled testing studies for Java programs}. In \bibinfo{booktitle}{\emph{Proceedings of the 2014 international symposium on software testing and analysis}}. \bibinfo{pages}{437--440}.
\newblock


\bibitem[Khanfir et~al\mbox{.}(2022)]%
        {khanfir2022codebert-nt}
\bibfield{author}{\bibinfo{person}{Ahmed Khanfir}, \bibinfo{person}{Matthieu Jimenez}, \bibinfo{person}{Mike Papadakis}, {and} \bibinfo{person}{Yves Le~Traon}.} \bibinfo{year}{2022}\natexlab{}.
\newblock \showarticletitle{Codebert-nt: code naturalness via codebert}. In \bibinfo{booktitle}{\emph{2022 IEEE 22nd International Conference on Software Quality, Reliability and Security (QRS)}}. IEEE, \bibinfo{pages}{936--947}.
\newblock


\bibitem[Kowsari et~al\mbox{.}(2019)]%
        {kowsari2019text}
\bibfield{author}{\bibinfo{person}{Kamran Kowsari}, \bibinfo{person}{Kiana Jafari~Meimandi}, \bibinfo{person}{Mojtaba Heidarysafa}, \bibinfo{person}{Sanjana Mendu}, \bibinfo{person}{Laura Barnes}, {and} \bibinfo{person}{Donald Brown}.} \bibinfo{year}{2019}\natexlab{}.
\newblock \showarticletitle{Text classification algorithms: A survey}.
\newblock \bibinfo{journal}{\emph{Information}} \bibinfo{volume}{10}, \bibinfo{number}{4} (\bibinfo{year}{2019}), \bibinfo{pages}{150}.
\newblock


\bibitem[Le et~al\mbox{.}(2022)]%
        {le2022coderl}
\bibfield{author}{\bibinfo{person}{Hung Le}, \bibinfo{person}{Yue Wang}, \bibinfo{person}{Akhilesh~Deepak Gotmare}, \bibinfo{person}{Silvio Savarese}, {and} \bibinfo{person}{Steven Chu~Hong Hoi}.} \bibinfo{year}{2022}\natexlab{}.
\newblock \showarticletitle{Coderl: Mastering code generation through pretrained models and deep reinforcement learning}.
\newblock \bibinfo{journal}{\emph{Advances in Neural Information Processing Systems}}  \bibinfo{volume}{35} (\bibinfo{year}{2022}), \bibinfo{pages}{21314--21328}.
\newblock


\bibitem[Li et~al\mbox{.}(2019)]%
        {li2019deepfl}
\bibfield{author}{\bibinfo{person}{Xia Li}, \bibinfo{person}{Wei Li}, \bibinfo{person}{Yuqun Zhang}, {and} \bibinfo{person}{Lingming Zhang}.} \bibinfo{year}{2019}\natexlab{}.
\newblock \showarticletitle{Deepfl: Integrating multiple fault diagnosis dimensions for deep fault localization}. In \bibinfo{booktitle}{\emph{Proceedings of the 28th ACM SIGSOFT international symposium on software testing and analysis}}. \bibinfo{pages}{169--180}.
\newblock


\bibitem[Loper and Bird(2002)]%
        {loper2002nltk}
\bibfield{author}{\bibinfo{person}{Edward Loper} {and} \bibinfo{person}{Steven Bird}.} \bibinfo{year}{2002}\natexlab{}.
\newblock \showarticletitle{NLTK: the Natural Language Toolkit}. In \bibinfo{booktitle}{\emph{Proceedings of the ACL-02 Workshop on Effective tools and methodologies for teaching natural language processing and computational linguistics-Volume 1}}. \bibinfo{pages}{63--70}.
\newblock


\bibitem[Ma et~al\mbox{.}(2022)]%
        {10.1145/3524842.3528456}
\bibfield{author}{\bibinfo{person}{Wei Ma}, \bibinfo{person}{Mengjie Zhao}, \bibinfo{person}{Ezekiel Soremekun}, \bibinfo{person}{Qiang Hu}, \bibinfo{person}{Jie~M. Zhang}, \bibinfo{person}{Mike Papadakis}, \bibinfo{person}{Maxime Cordy}, \bibinfo{person}{Xiaofei Xie}, {and} \bibinfo{person}{Yves~Le Traon}.} \bibinfo{year}{2022}\natexlab{}.
\newblock \showarticletitle{GraphCode2Vec: Generic Code Embedding via Lexical and Program Dependence Analyses}. In \bibinfo{booktitle}{\emph{Proceedings of the 19th International Conference on Mining Software Repositories}} (Pittsburgh, Pennsylvania) \emph{(\bibinfo{series}{MSR '22})}. \bibinfo{publisher}{Association for Computing Machinery}, \bibinfo{address}{New York, NY, USA}, \bibinfo{pages}{524–536}.
\newblock
\showISBNx{9781450393034}
\urldef\tempurl%
\url{https://doi.org/10.1145/3524842.3528456}
\showDOI{\tempurl}


\bibitem[Nijkamp et~al\mbox{.}(2022)]%
        {nijkamp2022codegen}
\bibfield{author}{\bibinfo{person}{Erik Nijkamp}, \bibinfo{person}{Bo Pang}, \bibinfo{person}{Hiroaki Hayashi}, \bibinfo{person}{Lifu Tu}, \bibinfo{person}{Huan Wang}, \bibinfo{person}{Yingbo Zhou}, \bibinfo{person}{Silvio Savarese}, {and} \bibinfo{person}{Caiming Xiong}.} \bibinfo{year}{2022}\natexlab{}.
\newblock \showarticletitle{Codegen: An open large language model for code with multi-turn program synthesis}.
\newblock \bibinfo{journal}{\emph{arXiv preprint arXiv:2203.13474}} (\bibinfo{year}{2022}).
\newblock


\bibitem[Olausson et~al\mbox{.}(2023)]%
        {olausson2023demystifying}
\bibfield{author}{\bibinfo{person}{Theo~X Olausson}, \bibinfo{person}{Jeevana~Priya Inala}, \bibinfo{person}{Chenglong Wang}, \bibinfo{person}{Jianfeng Gao}, {and} \bibinfo{person}{Armando Solar-Lezama}.} \bibinfo{year}{2023}\natexlab{}.
\newblock \showarticletitle{Demystifying GPT Self-Repair for Code Generation}.
\newblock \bibinfo{journal}{\emph{arXiv preprint arXiv:2306.09896}} (\bibinfo{year}{2023}).
\newblock


\bibitem[Patra and Pradel(2021)]%
        {mutate_motivation2}
\bibfield{author}{\bibinfo{person}{Jibesh Patra} {and} \bibinfo{person}{Michael Pradel}.} \bibinfo{year}{2021}\natexlab{}.
\newblock \showarticletitle{Semantic bug seeding: a learning-based approach for creating realistic bugs}. In \bibinfo{booktitle}{\emph{Proceedings of the 29th ACM Joint Meeting on European Software Engineering Conference and Symposium on the Foundations of Software Engineering}}. \bibinfo{pages}{906--918}.
\newblock


\bibitem[Pilkington(1996)]%
        {pilkington1996language}
\bibfield{author}{\bibinfo{person}{Adrian Pilkington}.} \bibinfo{year}{1996}\natexlab{}.
\newblock \showarticletitle{The Language Instinct: The New Science of Language and Mind: by Steven Pinker, 1994, The Penguin Press, London pp. 494, ISBN 0 713 99099 6 (hbk.)}.
\newblock \bibinfo{journal}{\emph{Language and Literature}} \bibinfo{volume}{5}, \bibinfo{number}{1} (\bibinfo{year}{1996}), \bibinfo{pages}{71--74}.
\newblock


\bibitem[Radford et~al\mbox{.}(2019)]%
        {radford2019language}
\bibfield{author}{\bibinfo{person}{Alec Radford}, \bibinfo{person}{Jeffrey Wu}, \bibinfo{person}{Rewon Child}, \bibinfo{person}{David Luan}, \bibinfo{person}{Dario Amodei}, \bibinfo{person}{Ilya Sutskever}, {et~al\mbox{.}}} \bibinfo{year}{2019}\natexlab{}.
\newblock \showarticletitle{Language models are unsupervised multitask learners}.
\newblock \bibinfo{journal}{\emph{OpenAI blog}} \bibinfo{volume}{1}, \bibinfo{number}{8} (\bibinfo{year}{2019}), \bibinfo{pages}{9}.
\newblock


\bibitem[Rahman et~al\mbox{.}(2019)]%
        {rahman2019natural_revisited}
\bibfield{author}{\bibinfo{person}{Musfiqur Rahman}, \bibinfo{person}{Dharani Palani}, {and} \bibinfo{person}{Peter~C Rigby}.} \bibinfo{year}{2019}\natexlab{}.
\newblock \showarticletitle{Natural software revisited}. In \bibinfo{booktitle}{\emph{2019 IEEE/ACM 41st International Conference on Software Engineering (ICSE)}}. IEEE, \bibinfo{pages}{37--48}.
\newblock


\bibitem[Ray et~al\mbox{.}(2016)]%
        {buggy_code_naturalness}
\bibfield{author}{\bibinfo{person}{Baishakhi Ray}, \bibinfo{person}{Vincent Hellendoorn}, \bibinfo{person}{Saheel Godhane}, \bibinfo{person}{Zhaopeng Tu}, \bibinfo{person}{Alberto Bacchelli}, {and} \bibinfo{person}{Premkumar Devanbu}.} \bibinfo{year}{2016}\natexlab{}.
\newblock \showarticletitle{On the" naturalness" of buggy code}. In \bibinfo{booktitle}{\emph{Proceedings of the 38th International Conference on Software Engineering}}. \bibinfo{pages}{428--439}.
\newblock


\bibitem[Raychev et~al\mbox{.}(2015)]%
        {raychev2015predicting}
\bibfield{author}{\bibinfo{person}{Veselin Raychev}, \bibinfo{person}{Martin Vechev}, {and} \bibinfo{person}{Andreas Krause}.} \bibinfo{year}{2015}\natexlab{}.
\newblock \showarticletitle{Predicting program properties from" big code"}.
\newblock \bibinfo{journal}{\emph{ACM SIGPLAN Notices}} \bibinfo{volume}{50}, \bibinfo{number}{1} (\bibinfo{year}{2015}), \bibinfo{pages}{111--124}.
\newblock


\bibitem[Raychev et~al\mbox{.}(2014)]%
        {2014Completion}
\bibfield{author}{\bibinfo{person}{Veselin Raychev}, \bibinfo{person}{Martin Vechev}, {and} \bibinfo{person}{Eran Yahav}.} \bibinfo{year}{2014}\natexlab{}.
\newblock \showarticletitle{Code completion with statistical language models}.
\newblock \bibinfo{journal}{\emph{Acm Sigplan Notices}} \bibinfo{volume}{49}, \bibinfo{number}{6} (\bibinfo{year}{2014}).
\newblock


\bibitem[Ren et~al\mbox{.}(2020)]%
        {Ren2020CodeBLEU}
\bibfield{author}{\bibinfo{person}{Shuo Ren}, \bibinfo{person}{Daya Guo}, \bibinfo{person}{Shuai Lu}, \bibinfo{person}{Long Zhou}, \bibinfo{person}{Shujie Liu}, \bibinfo{person}{Duyu Tang}, \bibinfo{person}{Neel Sundaresan}, \bibinfo{person}{Ming Zhou}, \bibinfo{person}{Ambrosio Blanco}, {and} \bibinfo{person}{Shuai Ma}.} \bibinfo{year}{2020}\natexlab{}.
\newblock \showarticletitle{CodeBLEU: a Method for Automatic Evaluation of Code Synthesis}.
\newblock \bibinfo{journal}{\emph{CoRR}}  \bibinfo{volume}{abs/2009.10297} (\bibinfo{year}{2020}).
\newblock


\bibitem[Roy et~al\mbox{.}(2021)]%
        {roy2021reassessing}
\bibfield{author}{\bibinfo{person}{Devjeet Roy}, \bibinfo{person}{Sarah Fakhoury}, {and} \bibinfo{person}{Venera Arnaoudova}.} \bibinfo{year}{2021}\natexlab{}.
\newblock \showarticletitle{Reassessing automatic evaluation metrics for code summarization tasks}. In \bibinfo{booktitle}{\emph{Proceedings of the 29th ACM Joint Meeting on European Software Engineering Conference and Symposium on the Foundations of Software Engineering}}. \bibinfo{pages}{1105--1116}.
\newblock


\bibitem[Roziere et~al\mbox{.}(2023)]%
        {codellama}
\bibfield{author}{\bibinfo{person}{Baptiste Roziere}, \bibinfo{person}{Jonas Gehring}, \bibinfo{person}{Fabian Gloeckle}, \bibinfo{person}{Sten Sootla}, \bibinfo{person}{Itai Gat}, \bibinfo{person}{Xiaoqing~Ellen Tan}, \bibinfo{person}{Yossi Adi}, \bibinfo{person}{Jingyu Liu}, \bibinfo{person}{Tal Remez}, \bibinfo{person}{J{\'e}r{\'e}my Rapin}, {et~al\mbox{.}}} \bibinfo{year}{2023}\natexlab{}.
\newblock \showarticletitle{Code llama: Open foundation models for code}.
\newblock \bibinfo{journal}{\emph{arXiv preprint arXiv:2308.12950}} (\bibinfo{year}{2023}).
\newblock


\bibitem[Suo et~al\mbox{.}(2024)]%
        {suo2024mlir}
\bibfield{author}{\bibinfo{person}{Chenyao Suo}, \bibinfo{person}{Junjie Chen}, \bibinfo{person}{Shuang Liu}, \bibinfo{person}{Jiajun Jiang}, \bibinfo{person}{Yingquan Zhao}, {and} \bibinfo{person}{Jianrong Wang}.} \bibinfo{year}{2024}\natexlab{}.
\newblock \showarticletitle{Fuzzing MLIR Compiler Infrastructure via Operation Dependency Analysis}. In \bibinfo{booktitle}{\emph{Proceedings of the 33nd ACM SIGSOFT International Symposium on Software Testing and Analysis}}.
\newblock


\bibitem[Tian and Chen(2023)]%
        {tian2023testcasedriven}
\bibfield{author}{\bibinfo{person}{Zhao Tian} {and} \bibinfo{person}{Junjie Chen}.} \bibinfo{year}{2023}\natexlab{}.
\newblock \bibinfo{title}{Test-Case-Driven Programming Understanding in Large Language Models for Better Code Generation}.
\newblock
\newblock
\showeprint[arxiv]{2309.16120}~[cs.SE]


\bibitem[Touvron et~al\mbox{.}(2023)]%
        {touvron2023llama}
\bibfield{author}{\bibinfo{person}{Hugo Touvron}, \bibinfo{person}{Thibaut Lavril}, \bibinfo{person}{Gautier Izacard}, \bibinfo{person}{Xavier Martinet}, \bibinfo{person}{Marie-Anne Lachaux}, \bibinfo{person}{Timoth{\'e}e Lacroix}, \bibinfo{person}{Baptiste Rozi{\`e}re}, \bibinfo{person}{Naman Goyal}, \bibinfo{person}{Eric Hambro}, \bibinfo{person}{Faisal Azhar}, {et~al\mbox{.}}} \bibinfo{year}{2023}\natexlab{}.
\newblock \showarticletitle{Llama: Open and efficient foundation language models}.
\newblock \bibinfo{journal}{\emph{arXiv preprint arXiv:2302.13971}} (\bibinfo{year}{2023}).
\newblock


\bibitem[Tu et~al\mbox{.}(2014)]%
        {tu2014localness}
\bibfield{author}{\bibinfo{person}{Zhaopeng Tu}, \bibinfo{person}{Zhendong Su}, {and} \bibinfo{person}{Premkumar Devanbu}.} \bibinfo{year}{2014}\natexlab{}.
\newblock \showarticletitle{On the localness of software}. In \bibinfo{booktitle}{\emph{Proceedings of the 22nd ACM SIGSOFT International Symposium on Foundations of Software Engineering}}. \bibinfo{pages}{269--280}.
\newblock


\bibitem[Vall{\'e}e-Rai et~al\mbox{.}(1999)]%
        {vallee1999soot}
\bibfield{author}{\bibinfo{person}{Raja Vall{\'e}e-Rai}, \bibinfo{person}{Phong Co}, \bibinfo{person}{Etienne Gagnon}, \bibinfo{person}{Laurie Hendren}, \bibinfo{person}{Patrick Lam}, {and} \bibinfo{person}{Vijay Sundaresan}.} \bibinfo{year}{1999}\natexlab{}.
\newblock \showarticletitle{Soot-a Java bytecode optimization framework}. In \bibinfo{booktitle}{\emph{Proceedings of the 1999 conference of the Centre for Advanced Studies on Collaborative research}}. \bibinfo{pages}{13}.
\newblock


\bibitem[Wang et~al\mbox{.}(2021a)]%
        {wang2021codet5}
\bibfield{author}{\bibinfo{person}{Yue Wang}, \bibinfo{person}{Weishi Wang}, \bibinfo{person}{Shafiq Joty}, {and} \bibinfo{person}{Steven~CH Hoi}.} \bibinfo{year}{2021}\natexlab{a}.
\newblock \showarticletitle{CodeT5: Identifier-aware Unified Pre-trained Encoder-Decoder Models for Code Understanding and Generation}. In \bibinfo{booktitle}{\emph{Proceedings of the 2021 Conference on Empirical Methods in Natural Language Processing}}. \bibinfo{pages}{8696--8708}.
\newblock


\bibitem[Wang et~al\mbox{.}(2021b)]%
        {wang2021prioritizing}
\bibfield{author}{\bibinfo{person}{Zan Wang}, \bibinfo{person}{Hanmo You}, \bibinfo{person}{Junjie Chen}, \bibinfo{person}{Yingyi Zhang}, \bibinfo{person}{Xuyuan Dong}, {and} \bibinfo{person}{Wenbin Zhang}.} \bibinfo{year}{2021}\natexlab{b}.
\newblock \showarticletitle{Prioritizing test inputs for deep neural networks via mutation analysis}. In \bibinfo{booktitle}{\emph{2021 IEEE/ACM 43rd International Conference on Software Engineering (ICSE)}}. IEEE, \bibinfo{pages}{397--409}.
\newblock


\bibitem[Woolson(2007)]%
        {woolson2007wilcoxon}
\bibfield{author}{\bibinfo{person}{Robert~F Woolson}.} \bibinfo{year}{2007}\natexlab{}.
\newblock \showarticletitle{Wilcoxon signed-rank test}.
\newblock \bibinfo{journal}{\emph{Wiley encyclopedia of clinical trials}} (\bibinfo{year}{2007}), \bibinfo{pages}{1--3}.
\newblock


\bibitem[Worster and Haines(2004)]%
        {mrr}
\bibfield{author}{\bibinfo{person}{Andrew Worster} {and} \bibinfo{person}{Ted Haines}.} \bibinfo{year}{2004}\natexlab{}.
\newblock \showarticletitle{Advanced statistics: understanding medical record review (MRR) studies}.
\newblock \bibinfo{journal}{\emph{Academic emergency medicine}} \bibinfo{volume}{11}, \bibinfo{number}{2} (\bibinfo{year}{2004}), \bibinfo{pages}{187--192}.
\newblock


\bibitem[Xu and Sheng(2024)]%
        {xu2024detecting}
\bibfield{author}{\bibinfo{person}{Zhenyu Xu} {and} \bibinfo{person}{Victor~S Sheng}.} \bibinfo{year}{2024}\natexlab{}.
\newblock \showarticletitle{Detecting AI-Generated Code Assignments Using Perplexity of Large Language Models}. In \bibinfo{booktitle}{\emph{Proceedings of the AAAI Conference on Artificial Intelligence}}, Vol.~\bibinfo{volume}{38}. \bibinfo{pages}{23155--23162}.
\newblock


\bibitem[Yan et~al\mbox{.}(2023)]%
        {yan2023coco}
\bibfield{author}{\bibinfo{person}{Ming Yan}, \bibinfo{person}{Junjie Chen}, \bibinfo{person}{Jie~M Zhang}, \bibinfo{person}{Xuejie Cao}, \bibinfo{person}{Chen Yang}, {and} \bibinfo{person}{Mark Harman}.} \bibinfo{year}{2023}\natexlab{}.
\newblock \showarticletitle{Coco: Testing code generation systems via concretized instructions}.
\newblock \bibinfo{journal}{\emph{arXiv preprint arXiv:2308.13319}} (\bibinfo{year}{2023}).
\newblock


\bibitem[Yan et~al\mbox{.}(2020)]%
        {jit}
\bibfield{author}{\bibinfo{person}{Meng Yan}, \bibinfo{person}{Xin Xia}, \bibinfo{person}{Yuanrui Fan}, \bibinfo{person}{Ahmed~E Hassan}, \bibinfo{person}{David Lo}, {and} \bibinfo{person}{Shanping Li}.} \bibinfo{year}{2020}\natexlab{}.
\newblock \showarticletitle{Just-in-time defect identification and localization: A two-phase framework}.
\newblock \bibinfo{journal}{\emph{IEEE Transactions on Software Engineering}} \bibinfo{volume}{48}, \bibinfo{number}{1} (\bibinfo{year}{2020}), \bibinfo{pages}{82--101}.
\newblock


\bibitem[Yang et~al\mbox{.}(2023)]%
        {yang2023silent}
\bibfield{author}{\bibinfo{person}{Chen Yang}, \bibinfo{person}{Junjie Chen}, \bibinfo{person}{Xingyu Fan}, \bibinfo{person}{Jiajun Jiang}, {and} \bibinfo{person}{Jun Sun}.} \bibinfo{year}{2023}\natexlab{}.
\newblock \showarticletitle{Silent Compiler Bug De-duplication via Three-Dimensional Analysis}. In \bibinfo{booktitle}{\emph{Proceedings of the 32nd ACM SIGSOFT International Symposium on Software Testing and Analysis}}. \bibinfo{pages}{677--689}.
\newblock


\bibitem[Yang et~al\mbox{.}(2024a)]%
        {chenyang_2024_12783666}
\bibfield{author}{\bibinfo{person}{Chen Yang}, \bibinfo{person}{Junjie Chen}, \bibinfo{person}{Jiajun Jiang}, {and} \bibinfo{person}{Yuliang Huang}.} \bibinfo{year}{2024}\natexlab{a}.
\newblock \bibinfo{booktitle}{\emph{Dependency-aware code naturalness}}.
\newblock
\urldef\tempurl%
\url{https://doi.org/10.5281/zenodo.12783666}
\showDOI{\tempurl}


\bibitem[Yang et~al\mbox{.}(2024b)]%
        {yang2024enhancing}
\bibfield{author}{\bibinfo{person}{Chen Yang}, \bibinfo{person}{Junjie Chen}, \bibinfo{person}{Bin Lin}, \bibinfo{person}{Jianyi Zhou}, {and} \bibinfo{person}{Ziqi Wang}.} \bibinfo{year}{2024}\natexlab{b}.
\newblock \showarticletitle{Enhancing LLM-based Test Generation for Hard-to-Cover Branches via Program Analysis}.
\newblock \bibinfo{journal}{\emph{arXiv preprint arXiv:2404.04966}} (\bibinfo{year}{2024}).
\newblock


\bibitem[Yang et~al\mbox{.}(2024c)]%
        {yang2024important}
\bibfield{author}{\bibinfo{person}{Guang Yang}, \bibinfo{person}{Yu Zhou}, \bibinfo{person}{Wenhua Yang}, \bibinfo{person}{Tao Yue}, \bibinfo{person}{Xiang Chen}, {and} \bibinfo{person}{Taolue Chen}.} \bibinfo{year}{2024}\natexlab{c}.
\newblock \showarticletitle{How important are good method names in neural code generation? a model robustness perspective}.
\newblock \bibinfo{journal}{\emph{ACM Transactions on Software Engineering and Methodology}} \bibinfo{volume}{33}, \bibinfo{number}{3} (\bibinfo{year}{2024}), \bibinfo{pages}{1--35}.
\newblock


\bibitem[Zan et~al\mbox{.}(2022a)]%
        {CERT}
\bibfield{author}{\bibinfo{person}{Daoguang Zan}, \bibinfo{person}{Bei Chen}, \bibinfo{person}{Dejian Yang}, \bibinfo{person}{Zeqi Lin}, \bibinfo{person}{Minsu Kim}, \bibinfo{person}{Bei Guan}, \bibinfo{person}{Yongji Wang}, \bibinfo{person}{Weizhu Chen}, {and} \bibinfo{person}{Jian-Guang Lou}.} \bibinfo{year}{2022}\natexlab{a}.
\newblock \showarticletitle{{CERT}: Continual Pre-training on Sketches for Library-oriented Code Generation}. In \bibinfo{booktitle}{\emph{The 2022 International Joint Conference on Artificial Intelligence}}.
\newblock


\bibitem[Zan et~al\mbox{.}(2022b)]%
        {msra_survey}
\bibfield{author}{\bibinfo{person}{Daoguang Zan}, \bibinfo{person}{Bei Chen}, \bibinfo{person}{Fengji Zhang}, \bibinfo{person}{Dianjie Lu}, \bibinfo{person}{Bingchao Wu}, \bibinfo{person}{Bei Guan}, \bibinfo{person}{Yongji Wang}, {and} \bibinfo{person}{Jian-Guang Lou}.} \bibinfo{year}{2022}\natexlab{b}.
\newblock \showarticletitle{Large language models meet nl2code: A survey}.
\newblock \bibinfo{journal}{\emph{arXiv preprint arXiv:2212.09420}} (\bibinfo{year}{2022}).
\newblock


\bibitem[Zhang et~al\mbox{.}(2023)]%
        {zhang2023planning}
\bibfield{author}{\bibinfo{person}{Shun Zhang}, \bibinfo{person}{Zhenfang Chen}, \bibinfo{person}{Yikang Shen}, \bibinfo{person}{Mingyu Ding}, \bibinfo{person}{Joshua~B Tenenbaum}, {and} \bibinfo{person}{Chuang Gan}.} \bibinfo{year}{2023}\natexlab{}.
\newblock \showarticletitle{Planning with large language models for code generation}.
\newblock \bibinfo{journal}{\emph{arXiv preprint arXiv:2303.05510}} (\bibinfo{year}{2023}).
\newblock


\bibitem[Zheng et~al\mbox{.}(2023)]%
        {zheng2023codegeex}
\bibfield{author}{\bibinfo{person}{Qinkai Zheng}, \bibinfo{person}{Xiao Xia}, \bibinfo{person}{Xu Zou}, \bibinfo{person}{Yuxiao Dong}, \bibinfo{person}{Shan Wang}, \bibinfo{person}{Yufei Xue}, \bibinfo{person}{Zihan Wang}, \bibinfo{person}{Lei Shen}, \bibinfo{person}{Andi Wang}, \bibinfo{person}{Yang Li}, {et~al\mbox{.}}} \bibinfo{year}{2023}\natexlab{}.
\newblock \showarticletitle{Codegeex: A pre-trained model for code generation with multilingual evaluations on humaneval-x}.
\newblock \bibinfo{journal}{\emph{arXiv preprint arXiv:2303.17568}} (\bibinfo{year}{2023}).
\newblock


\end{thebibliography}

\end{document}